\documentclass[a4paper,10pt]{article}

\pdfoutput=1
\usepackage{jheppub1}
\usepackage{array}
\usepackage{bbold}
\usepackage{multirow}
\usepackage{amsmath}
\usepackage{makecell}
\usepackage{braket}
\usepackage{enumerate}
  {\left.\begin{aligned}}
  {\end{aligned}\right\rbrace}

\definecolor{green}{rgb}{0.1,0.8,0.2}

\usepackage{cancel}
\usepackage{amsmath,amssymb} 
\usepackage{graphicx} 
\usepackage[dvipsnames,x11names]{xcolor} 

\usepackage{hyperref}
\usepackage{cleveref}
\usepackage{float}
\usepackage{subcaption}
\usepackage{tikz, pgfplots}
\usetikzlibrary{decorations}
\usepackage{caption}
\usetikzlibrary[positioning]
\usetikzlibrary{snakes}

\makeatletter
\newcommand{\footnoteref}[1]{\protected@xdef\@thefnmark{\ref{#1}}\@footnotemark}
\makeatother

\begin{document}

	\preprint{}
	\title{Moving Interfaces and two-dimensional Black Holes}
	\author[a]{Parthajit Biswas}
        \affiliation[a]{Department of Physics, Ramakrishna Mission Vivekananda Educational and Research Institute,\\
        Belur Math, Howrah 711202, India.}
     \author[b]{Suchetan Das}
     \affiliation[b]{Center for High Energy Physics,\\
Indian Institute of Science, Bangalore 560012, India}
     \author[c]{Anirban Dinda}
     \affiliation[c]{School of Physical Sciences, Indian Association for Cultivation of Science,\\
Kolkata 700032, West Bengal, India}
	\emailAdd{parthajitbiswas8@gmail.com, suchetan1993@gmail.com, dindaanirban@gmail.com}

	\abstract{Conformal field theories can exchange energy through a boundary interface. Imposing conformal boundary conditions for static interfaces implies energy conservation at the interface. Recently, the reflective and transmitive properties of such static conformal interfaces have been studied in two dimensions by scattering matter at the interface impurity. In this note, we generalize this to the case of dynamic interfaces. Motivated by the connections between the moving mirror and the black hole, we choose a particular profile for the dynamical interface. We show that a part of the total energy of each side will be lost in the interface. In other words, a time-dependent interface can accumulate or absorb energy. While, in general, the interface follows a time-like trajectory, one can take a particular limit of a profile parameter($\beta$), such that the interface approaches a null line asymptotically$(\beta\rightarrow 0)$. In this limit, we show that for a class of boundary conditions, the interface behaves like a semipermeable membrane - it behaves like a (partially) reflecting mirror from one side and is (partially) transparent from the other side. We also consider another set of conformal boundary conditions for which, in the null line limit, the interface mimics the properties expected of a horizon. In this case, we devise a scattering experiment, where (zero-point subtracted) energy from one CFT is fully transmitted to the other CFT, while from the other CFT, energy can neither be transmitted nor reflected, i.e., it gets lost in the interface. This boundary condition is also responsible for the thermal energy spectrum which mimics Hawking radiation. This is analogous to the black hole where the horizon plays the role of a one-sided `membrane', which accumulates all the interior degrees of freedom and radiates thermally in the presence of quantum fluctuation. Stimulated by this observation, we comment on some plausible construction of wormhole analogues. }


\maketitle

\section{Introduction} \label{sec:intro}
Boundary conditions play a crucial role in constraining dynamics in the real world. In particular, for studying non-equilibrium physics, the role of boundary condition is extremely important, e.g., thermalization from quench \cite{Calabrese:2016xau}. An interesting question in this context is whether we can mimic thermal-like behavior by imposing suitable boundary conditions. As was first shown in \cite{Davies:1976hi,Birrell:1982ix,PhysRevD.36.2327}, a real-time dynamical mirror indeed captures certain thermal features which eventually mimic black hole physics \cite{Hawking:1975vcx}. The setup involves a quantum field on 2$D$ flat spacetime with a perfectly reflecting boundary. Particles are created when the moving mirror undergoes an acceleration. If the mirror trajectory is suitably chosen, the average particle flux can be identified as the average particle production in the background of a collapsing black hole. Considering moving mirror as a boundary profile in the context of boundary conformal field theory (BCFT)\cite{Cardy:1984bb}, Akal et al. \cite{Akal:2020twv,Akal:2021foz} have computed time evolution of entanglement entropy for different moving mirror boundary profiles, and in particular for the set-up which mimics black hole formation and evaporation they have found an ideal Page curve\footnote{ A different entanglement measure in the moving mirror setup can be found in \cite{Kawabata:2021hac}.}. This interesting analogy between a dynamical boundary and black hole physics still lacks a formal identification or definition of the horizon and beyond horizon physics. A natural generalization of this program is to study different boundary conditions in such dynamical mirror profiles for which two different theories can interact along the interface\footnote{To the best of our knowledge, dynamical (or moving) interface has not been studied in the literature. However, \cite{Good:2021asq} (and references therein) addressed semi reflecting moving mirror in other context.}. 

Such interfaces allow the two CFTs to interact, thereby allowing the exchange of energy between them. The fraction of energy transport across the interface is characterized by \textit{transport coeffecients}\cite{Quella:2006de,Meineri:2019ycm}. The proportion of energy that is transmitted through the interface is called \textit{transmission coefficient}(${\cal T}$), and the proportion of energy that is bounced back from the interface is called \textit{reflection coefficient}(${\cal R}$). ${\cal T}$ and ${\cal R}$ are defined through the expectation values of the stress tensor \citep{Meineri:2019ycm}; they depend on the details of the CFTs and also on the boundary conditions used to consistently glue CFTs along the interface.

In \citep{Meineri:2019ycm}, the authors have studied transport coefficients in great detail by considering two $2D$ CFTs, $CFT_{L}$ and $CFT_{R}$ glued along a \textit{static} interface. The idea was to prepare an excitation far from the interface and let it collide with the interface and then compute ${\cal T}$ and ${\cal R}$ by computing the fraction of energy that has been transmitted and reflected by the interface. In particular, they have used a boundary condition known as \textit{conformal boundary condition}, which ensures that the amount of energy flowing towards the interface is the same as the amount of energy that is flowing away from the interface; in other words, their boundary condition ensures that the interface neither absorb nor radiate any energy. Obviously, they get transmission and reflection coefficients which satisfy ${\cal T}+{\cal R}=1$.


%
In this note, we generalize their setup to the case when the interface is dynamical. 
We then compute the transmission and reflection coefficients in this setup and discuss some probable connections with black holes.
%
\begin{figure}[h]
\centering
\begin{tikzpicture}[scale=0.7]
\draw[thick] (0,-2)--(0,2)--(6,2)--(6,-2)--(0,-2);
\filldraw[fill=green!40, thick] (1,2) parabola (3,-2)--(6,-2)--(6,2)--(1,2); 
\filldraw[fill=yellow!20, thick] (1,2) parabola (3,-2)--(0,-2)--(0,2)--(1,2); 
\draw[ultra thick,blue] (1,2) parabola (3,-2);
\draw [dashed, ->] (3,-2.5) -- (3,2.5) node at (3,2.7) {$ t$};
\draw [dashed] (-0.5,0) -- (0.8,0) ;
\draw [dashed] (2.2,0) -- (3.9,0) ;
\draw [dashed,->] (5.1,0) -- (6.5,0) node[right] {$x$};
\draw node at (1.5,0) {CFT$_{L}$};
\draw node at (4.5,0) {CFT$_R$};
\draw[->, very thick] (7.4,0)--(8.2,0);
\draw[thick] (9,-2)--(9,2)--(15,2)--(15,-2)--(9,-2);
\filldraw[fill=green!40, thick] (15,2)--(12,2)--(12,-2)--(15,-2)--(15,2); 
\filldraw[fill=yellow!20, thick] (12,2) -- (9,2)--(9,-2)--(12,-2)--(12,2); 
\draw[blue, ultra thick] (12,-2)--(12,2);
\draw [dashed, ->] (12,-2.5) -- (12,2.5) node at (12,2.8) {$ \tilde{t}$};
\draw [dashed] (8.5,0) -- (9.8,0) ;
\draw [dashed] (11.2,0) -- (12.9,0) ;
\draw [dashed,->] (14.2,0) -- (15.5,0) node[right] {$\tilde{x}$};
\draw node at (10.5,0) {CFT$_L$};
\draw node at (13.5,0) {CFT$_R$};
\end{tikzpicture}
\caption{Static interface from dynamical interface using conformal mapping.}
\label{fig:mesh1}
\end{figure}
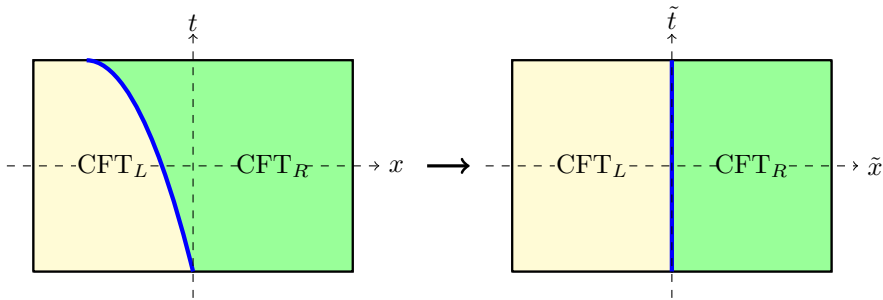
%
We will use the conformal map method \cite{Akal:2020twv,Akal:2021foz} to map the dynamical interface to a static interface (see fig.(\ref{fig:mesh1}) for a schematic representation). After going to the static interface, we will use conformal boundary condition of the product CFTs $CFT_{L} \otimes \overline{CFT}_{R}$\citep{Meineri:2019ycm}. The crucial feature of this conformal map is that the left-moving and right-moving energy flux are mapped differently. Even though the net energy flux in both sides will be the same due to this boundary condition, the moving interface profile itself contributes to the energy. So, in our case, we will not get ${\cal T}+{\cal R}=1$. The value of ${\cal T}+{\cal R}$ depends on the details of the interface profile. However, in this note, we will only consider one particular interface profile, namely \textit{escaping-mirror profile}, which is known to model Hawking radiation from a black hole formed by a collapsing null shell. In the static case, reflection and transmission coefficients surprisingly do not depend on any details of the state where we measure energy in the collider experiment. However, in our dynamical case, these transport coefficients indeed depend on the profile of the interface as well as the position of incoming excitations where the initial state is prepared. This is a clear manifestation of the breaking of conformal symmetries of the moving interface. In a particular limit of the parameter of the interface profile ($\beta \rightarrow 0$), we could obtain $\mathcal{R} \rightarrow 0 $ or $\mathcal{T} \rightarrow 1 $. This feature will motivate us to study another set of boundary conditions which is a conformal boundary condition of product CFTs $CFT_{L}\otimes CFT_{R}$. We can embed it in the standard two-sided ICFT picture by exchanging chiral and anti-chiral sectors of one CFT. In this setup, we modify the scattering experiment discussed in \cite{Meineri:2019ycm} to make it consistent with this exchange.
This setting is natural for such gluing conditions. For a particular solution to this gluing condition (\ref{bc3}), we can perform the modified scattering experiment for which the interface behaves like a one-sided transmitter with no reflection from either side at $\beta \rightarrow 0$ when the interface coincides with null coordinate. This is one of the defining properties of a classical horizon that is realized in our moving interface setup. One can also argue that the thermal spectrum of Hawking radiation could be obtained using this same boundary condition in a similar manner to moving mirror setup. The same gluing condition could also be realized as an ICFT if we reverse the time direction of one copy. In this case, we could also perform a scattering experiment where the past and future (null) infinity of one side will be reversed. In this setting, we will show that the two CFTs subjected to gluing condition (\ref{bc3}) behave like two exteriors of an eternal black hole. We make this connection more precise by obtaining Hawking radiation in each side and by identifying the vacuum as a thermofield double(TFD) state with respect to future observers.

The rest of the note is organized as follows: In \S\ref{sec:static_review}, we will briefly review CFTs in the presence of a static interface; we will discuss computations of transmission and reflection coefficients following a collider experiment. In \S\ref{ref:moving1}, we will generalize the computation of transmission and reflection coefficients for a dynamical interface by using the \textit{conformal map method} \cite{Akal:2021foz}. 
In the limit, when the interface becomes null, we will show that the interface behaves like a \textit{semipermeable} membrane. 
In \S\ref{diffbdy}, we will consider a slightly different boundary condition and will compute transmission and reflection coefficients. We will show that the interface again behaves like a \textit{semipermiable} membrane but, contrary to the previous case, gives thermal Hawking flux, thus providing connections with black hole physics. In \S\ref{sec:Horizon}, we will consider the later boundary condition and will do a parity and time-reversal transformation to get a natural ICFT interpretation, keeping the boundary condition intact and will interpret these two CFTs as two exteriors of an eternal black hole. We will end with discussions and future directions in \S\ref{sec:discussion}. Various technical details will be presented in appendices \ref{A1} and \ref{app:B}.

\section{CFT with static interface: a brief review}\label{sec:static_review}
In this section, we will briefly review the computations of transmission and reflection coefficients in the presence of a static interface as described in\citep{Meineri:2019ycm}.

\subsection{Interface gluing Conditions}
Consider two Lorentzian CFTs, $CFT_L$, and $CFT_R$, separated by an interface (defect/impurity) placed at $x=0$. The two CFTs, in general, may have different central charges. On the interface, the stress tensors of the two CFTs are subject to the following conformal gluing condition\cite{Quella:2006de}.
\begin{equation}\label{standard icft}
T_{L}(u)+\bar{T}_{R}(v)=\bar{T}_{L}(v)+T_{R}(u)\Big{|}_{u=v},\quad u \equiv t-x, v\equiv t+x
\end{equation}
These gluing conditions at the interface capture the coupling of the two CFTs via some boundary interactions localized on the interface. 
Two extreme solutions to the above equation correspond to having no coupling between the $CFT_L$ and $CFT_R$  or complete transparency between them.  We refer to these as `factorizing gluing condition' and `transparent gluing condition', respectively. 
\begin{itemize}
\item{ Factorizing gluing condition}
\begin{equation}
T_{L}(u)=\bar{T}_{L}(v) \hspace{.5cm} T_{R}(u)=\bar{T}_{R}({v})\,.
\end{equation}
This results in two decoupled BCFTs.
\item Transparent gluing condition 
\begin{equation}
T_{L}(u)=T_{R}(u) \hspace{.5cm} \bar{T}_{L}(v)=\bar{T}_{R}(v).
\end{equation}
With this condition, the left-moving and right-moving stress tensors become continuous across the interface, and the theory becomes independent of the location of the interface.
\end{itemize}
%
Between these two extreme cases, one may have interfaces with partial reflectivity, which can be encoded in nonvanishing reflection and transmission coefficients.
 
The two-point functions of the stress tensors on the same side are completely fixed by conformal symmetry.
\begin{equation}
\begin{split}
\braket{T_{L}(u_{1})T_{L}(u_{2})}_{I}=\frac{c_{L}/2}{(u_{1}-u_{2})^{4}} \hspace{.5cm} \braket{T_{R}(u_{1})T_{R}(u_{2})}_{I}=\frac{c_{R}/2}{(u_{1}-u_{2})^{4}}
\end{split}
\end{equation}
\begin{equation}
\begin{split}
\braket{\bar{T}_{L}(v_{1})\bar{T}_{L}(v_{2})}_{I}=\frac{c_{L}/2}{(v_{1}-v_{2})^{4}} \hspace{.5cm} \braket{\bar{T}_{R}(v_{1})\bar{T}_{R}(v_{2}}_{I}=\frac{c_{R}/2}{(v_{1}-v_{2})^{4}}
\end{split}
\end{equation}
While computing the left-right stress tensor correlations, new coefficients that encode the coupling of the two theories will appear. 
\begin{equation}\label{eq:cLR}
\begin{split}
\braket{T_{L}(u_{1})T_{R}(u_{2})}_{I}=\frac{c_{LR}/2}{(u_{1}-u_{2})^{4}} \hspace{.5cm} \braket{\bar{T}_{L}(v_1)\bar{T}_{R}(v_{2})}_{I}=\frac{c_{LR}/2}{(v_{1}-v_{2})^{4}}
\end{split}
\end{equation}
\begin{equation}
\begin{split}
\braket{T_{L}(u)\bar{T}_{L}(v)}_{I}=\frac{(c_{L}-c_{LR})/2}{(u+v)^{4}} \hspace{.5cm} \braket{T_{R}(u)\bar{T}_{R}(v)}_{I}=\frac{(c_{R}-c_{LR})/2}{(u+v)^{4}}
\end{split}
\end{equation}

Here, $c_{L}$ and $c_{R}$ are the central charges of the two CFTs, respectively. $c_{LR}$ is the coefficient of the two-point function of the stress tensors across a static interface \cite{Meineri:2019ycm}. In this note, we will be interested in parity invariant CFTs for which $c_{L}=\bar{c}_{L}$, $c_{R}=\bar{c}_{R}$ and $c_{LR}=\bar{c}_{LR}$.

\subsection{Collider Experiment}\label{Collider Experiment}

We now briefly review the collider experiment set-up introduced in\cite{Meineri:2019ycm}.  The idea is to scatter conformal matter in one theory towards the interface and compute the amount of energy flux obtained after reflection and transmission at the interface. This flux is given by the ANEC operators \cite{Hofman:2008ar}, defined below:
%
\begin{equation}
\begin{split}
\mathcal{E} = -\frac{1}{2\pi} \int^{\infty}_{-\infty}   du T(u) \hspace{.5cm} \bar{\mathcal{E}} = -\frac{1}{2\pi} \int^{\infty}_{-\infty}   dv \bar{T}(v)   
\end{split}
\end{equation}

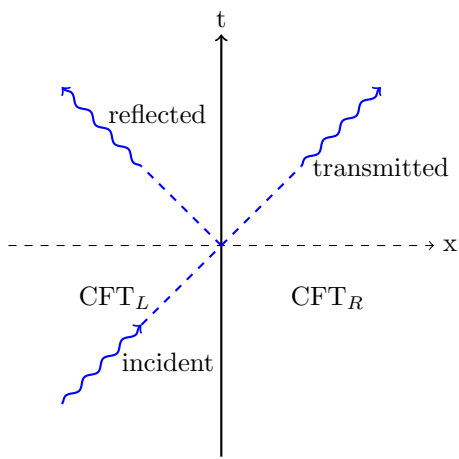
\begin{figure}[H]
\begin{center}
\begin{tikzpicture}[scale=0.7]
\draw[->,dashed] (-1,0) -- (7,0);
\draw[->, thick] (3,-4)--(3,4);
   \draw node at (5,-1) {CFT$_R$};
    \draw node at (1,-1) {CFT$_L$};
    \draw [dashed, blue, thick] (1.5,-1.5)--(3,0);
    \draw [dashed, blue, thick] (3,0)--(1.5,1.5);
    \draw [dashed, blue, thick] (3,0)--(4.5,1.5);
    \draw [->,decorate,decoration={snake,amplitude=.4mm}, blue, thick](0,-3) -- (1.5,-1.5);
    \draw [->,decorate,decoration={snake,amplitude=.4mm}, blue, thick](1.5,1.5) -- (0,3);
    \draw [->,decorate,decoration={snake,amplitude=.4mm}, blue, thick](4.5,1.5) -- (6,3);
    \draw node at (1.8,2.5) {reflected};
    \draw node at (2,-2.2) {incident};
    \draw node at (6,1.5) {transmitted};
    \draw node at (7.3,0) {x};
    \draw node at (3,4.3) {t};
    \end{tikzpicture}
\end{center}
 \caption{The interface which separates two CFTs, is placed at $x=0$}
    \label{fig:scattering}
\end{figure}

These compute the total energy flux at infinity towards the right or the left, respectively. In ICFT, the two independent components of the stress tensor move freely in the bulk but interact through the interface (see fig \ref{fig:scattering} for schematic representation). Using these ANEC operators, one can define  the transmission and reflection coefficients for energy flux across an interface as follows 
\begin{equation}
\begin{split}
\mathcal{T}= \frac{\text{transmitted energy}}{\text{incident energy}} \hspace{.5cm} \mathcal{R}= \frac{\text{reflected  energy}}{\text{incident energy}} 
\end{split}
\end{equation}

Energy conservation implies $\mathcal{T}+\mathcal{R}=1$, and positivity of the total energy transmitted and reflected leads to $\mathcal{T} \geq 0$ and $\mathcal{R} \geq 0$. A priori, one expects that $\mathcal{T}$ and $\mathcal{R}$ would depend on the details of the states. But remarkably, one can show that in a generic CFT, they are completely independent of the details of the incoming excitation. More precisely, one can show that they are entirely determined by the central charges of the two CFTs $c_{L},c_{R}$ and $c_{LR}$ \eqref{eq:cLR}. This is an important universal feature of a static interface\cite{Meineri:2019ycm}.

Being a scale-invariant theory, there are no well-defined asymptotic states in CFT. We follow the procedure described in \cite{Meineri:2019ycm} to create scattering experiments. One can define a one-parameter family of states:
\begin{equation}
\begin{split}
|\mathcal{O}_{L},D\bigr{>}_{I}= \int d^{2}x  f(u)f(v+D)\mathcal{O}_{L}(u,v)|O\bigr{>}_{I}
\end{split}
\end{equation}
$f(u)$ is some appropriate square integrable wave packet. 
\begin{equation}
\begin{split}
\int_{-\infty}^{\infty} |f(u)| du =1,   \hspace{.5cm} f(u)=0 \hspace{.1cm} \text{if} \hspace{.3cm} |u| > l
\end{split}
\end{equation}
where the support of the wave packet is $u\in (-l,l)$. 

Here, $D$ signifies how far the state is from the interface. We have introduced this parameter to set up scattering-like experiments. As $D\rightarrow\infty$, the effect of the boundary condition on the interface will vanish. In this way, we can define asymptotic states in the collider experiment. The normalization of the states reads as 
\begin{equation}
\begin{split}
\lim_{D \to \infty}\bigr{<}\mathcal{O}_{L},D|\mathcal{O}_{L},D\bigr{>}_{I}=\bigr{<}\mathcal{O}_{L},D|\mathcal{O}_{L},D\bigr{>} 
\end{split}
\end{equation}
As ${D \to \infty}$ limit prepares a state with no interface, we can drop the index $I$. 
Having defined the states, we are going to define the observables 
\begin{equation}\label{Transport defn}
\begin{split}
\mathcal{T}_{L }&= \lim_{D\to\infty} \frac{\braket{O_{L},D|\mathcal{E}_{R}|O_{L},D}_{I}}{\braket{{O_{L},D|\mathcal{E}_{L}|O_{L},D}}} \\ 
\mathcal{R}_{L }&= \lim_{D\to\infty} \frac{\braket{O_{L},D|\bar{\mathcal{E}}_{L}|O_{L},D}_{I}- \braket{O_{L},D|\bar{\mathcal{E}}_{L}|O_{L},D}}{\braket{{O_{L},D|\mathcal{E}_{L}|O_{L},D}}}
 \\ 
\mathcal{T}_{R }&= \lim_{D\to\infty} \frac{\braket{\bar{O}_{R},D|\bar{\mathcal{E}}_{L}|\bar{O}_{R},D}_{I}}{\braket{{\bar{O}_{R},D|\bar{\mathcal{E}}_{R}|\bar{O}_{R},D}}}
\ \\ 
\mathcal{R}_{R }&= \lim_{D\to\infty} \frac{\braket{\bar{O}_{R},D|\mathcal{E}_{R}|\bar{O}_{R},D}_{I}- \braket{\bar{O}_{R},D|\mathcal{E}_{R}|\bar{O}_{R},D}}{\braket{{\bar{O}_{R},D|\bar{\mathcal{E}}_{R}|\bar{O}_{R},D}}}
\end{split}
\end{equation}
The transmission coefficients for the static interface have the following expressions \cite{Meineri:2019ycm}
\begin{equation}
\begin{split}
\mathcal{T}_{L}=\frac{c_{LR}}{c_{L}}\,,\hspace{.5cm} \mathcal{T}_{R}=\frac{c_{LR}}{c_{R}} \,.
\end{split}
\end{equation}
%
For the two special solutions of the gluing condition, the transmission coefficients are  
\begin{equation}
\begin{split}
\text{Totally reflective} : \mathcal{T}_{L}=\mathcal{T}_{R}=0\\
\text{Totally transmissive} : \mathcal{T}_{L}=\mathcal{T}_{R}=1
\end{split}
\end{equation}

One can show the following bounds from the energy conservation and positivity of the total energy transmitted and reflected.  
\begin{equation}
\begin{split}
0<c_{LR} < \text{min}(c_{R},c_{L}) , \hspace{.2cm} 0<\mathcal{T}_{L} < \text{min} (1,\frac{c_{R}}{c_{L}}) , \hspace{.2cm} 0<\mathcal{T}_{R} < \text{min} (1,\frac{c_{L}}{c_{R}})
\end{split}
\end{equation}

\section{CFT with moving interface}\label{sec:moving_interface}

\subsection{Set up for collider experiment}\label{ref:moving1}
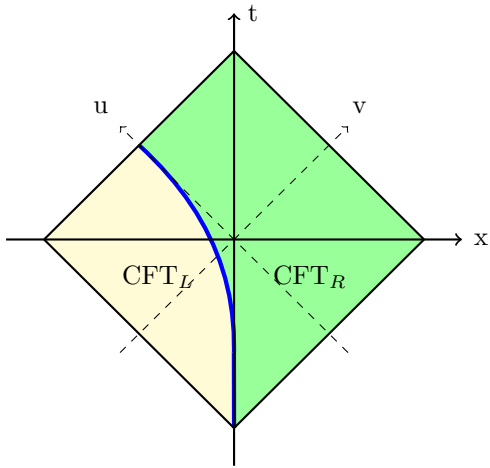
\begin{figure}[H]
    \centering
    \begin{tikzpicture}[scale=.5]
    \fill[fill=green!40](-5,0)--(0,5)--(5,0)--(0,-5)--(-5,0);
    \fill[fill=yellow!20,] (-.23,-4.77) to[bend right] (-2.5,2.5)--(-5,0)--(-.23,-4.77);
    \draw[ultra thick,blue] (-.23,-4.77) to[bend right] (-2.5,2.5);
    \fill[fill=yellow!20,] (0,-3)--(0,-5)--(-2,-3);
    \draw[color=blue, ultra thick] (-0.004,-3)--(0,-5);
    \draw [dashed, ->] (-3,-3) -- (3,3) node at (3.3,3.5) {v};
    \draw [dashed, ->] (3,-3) -- (-3,3) node at (-3.5,3.5) {u} ;
    \draw [thick, ->] (-6,0) -- (6,0) node at (6.5,0) {x};
    \draw [thick, ->] (0,-6) -- (0,6) node at (.5,6) {t};
    \draw[thick](-5,0)--(0,5)--(5,0)--(0,-5)--(-5,0);
    \draw node at (2,-1) {CFT$_R$};
    \draw node at (-2,-1) {CFT$_L$};
    \end{tikzpicture}
\caption{The moving interface profile: the blue line presents the motion of the mirror}
\label{fig:dynamic_interface}
\end{figure}

{\it \underline{Moving Interfaces}}: In this section, we generalize the discussion of the previous section to the case when the defect is moving along a time-like curve $x=Z(t)$, which in terms of the light cone coordinates ($u, v$) can be expressed as $v=p(u)$ (see figure \ref{fig:dynamic_interface}). The choice of the function $p(u)$ determines different trajectories of the interface. 

%
\begin{figure}
    \centering
   \begin{tikzpicture}[scale=.5]
    \draw[thin](-5,0)--(0,5)--(5,0)--(0,-5)--(-5,0);
    \filldraw[color=red!60, fill=red!30, very thick](-5,0)--(0,5)--(2.5,2.5)--(-2.5,-2.5)--(-5,0);
       \filldraw[color=black!6, fill=yellow!20, very thick](0,0)--(-2.5,-2.5)--(0,-5)--cycle;
       \filldraw[color=green!6, fill=green!40](0,0)--(2.5,2.5)--(5,0)--(0,-5)--(0,0);
       \draw [dashed, ->] (-3,-3) -- (3,3) node at (3.3,3.5) {$\tilde{v}$};
        \draw [dashed, ->] (3,-3) -- (-3,3) node at (-3.5,3.5) {$\tilde{u}$} ;
        \draw [thin, ->] (-6,0) -- (6,0) node at (6.5,0) {$\tilde{x}$};
        \draw [thin, ->] (0,-6) -- (0,6) node at (.5,6) {$\tilde{t}$};
        \draw[ultra thick, blue](0,-5)--(0,0);
        \draw node at (2,-1) {CFT$_R$};
        \draw node at (-1,-3) {CFT$_L$};
   \end{tikzpicture}
    \caption{After the transformation, the interface gets mapped to the blue solid line. The red-shaded area is not accessible \cite{Akal:2021foz}.}
    \label{fig:static_interface}
\end{figure}
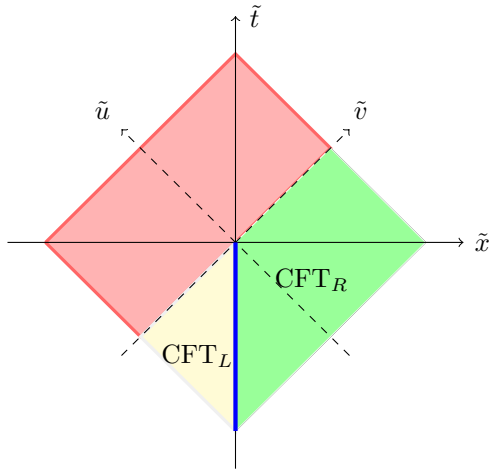

 Following \cite{Akal:2020twv}, one can map the dynamical interface problem to the static one (see figure \ref{fig:static_interface}) using the following conformal transformations:
\begin{align}\label{map}
    \tilde{v}=v, \; \tilde{u}=p(u), \; \text{where} \; \tilde{u} = \tilde{t}-\tilde{x}, \tilde{v}=\tilde{t}+\tilde{x}.
\end{align}

In the new ($\tilde{u},\tilde{v}$) coordinates, the position of the interface will be at $\Tilde{x}=0$. 
 In this note, we will only consider the escaping interface profile, which has the following form:
\begin{align}\label{escaping profile}
    p(u) = -\beta\log\left(1+e^{-\frac{u}{\beta}}\right), \; \beta \geq 0
\end{align}
%

In the $u \rightarrow \infty$, $p(u) \rightarrow -\beta e^{-\frac{u}{\beta}}$. On the other hand, if we take $u \rightarrow -\infty$, the profile becomes $p(u) \rightarrow u$. This essentially states that in the far past, the interface was static at $x=0$, and at very late time, it becomes null (see figure \ref{fig:dynamic_interface}).

Using the map (\ref{map}), one can compute the vacuum energy $\langle T_{uu}(u)\rangle|_{I}$ in the presence of the interface. This has the following form
\begin{align}\label{1}
    \langle T^{L}_{uu}(u)\rangle|_{I}= \frac{c_{L}}{48\pi\beta^{2}}\left(1-\frac{1}{(1+e^{\frac{u}{\beta}})^{2}}\right); \; \langle T^{R}_{uu}(u)\rangle|_{I} = \frac{c_{R}}{48\pi\beta^{2}}\left(1-\frac{1}{(1+e^{\frac{u}{\beta}})^{2}}\right)
\end{align}
In the asymptotic limit $u\rightarrow \infty$, this takes the form:
\begin{align}\label{2}
  \langle T^{L}_{uu}(u)\rangle|_{I} \approx  \frac{c_{L}}{48\pi\beta^{2}}; \; \langle T^{R}_{uu}(u)\rangle|_{I} \approx  \frac{c_{R}}{48\pi\beta^{2}} 
\end{align}
 While in the $u\rightarrow -\infty$ limit, we get
 \begin{align}\label{3}
 T^{L}_{uu} = T^{R}_{uu} =0
 \end{align}
 Interestingly, as is clear from the expressions for the stress tensors (\ref{1}), the asymptotic limits in $u$ can also be approached instead by taking a limit in $\beta$, keeping $u$ fixed and positive. 
%
If we consider $\beta\rightarrow 0$ for a fixed positive $u$ we recover (\ref{2}), while if we instead consider $\beta \rightarrow \infty$ for a fixed $u$, then we recover (\ref{3}).  These late-time features are very crucial for some of our later results presented in the next section. \\\\
\textit{\underline{Radiative ICFT}}: In the same spirit as \cite{Akal:2021foz}, we define `Radiative ICFTs' by mapping the standard ICFT boundary condition imposed on the static interface (\ref{standard icft}) to the moving interface \footnote{Here we subtracted the vacuum energy part of the stress tensor. In other words here $T \rightarrow T-\langle T \rangle_{vac}$.}
\begin{align}\label{sbc1}
(p'(u))^{-2}T_{R}(u)-\bar{T}_{R}(v)= (p'(u))^{-2}T_{L}(u)-\bar{T}_{L}(v) \; \text{at} \; v=p(u)
\end{align}
In the static frame, this reduces to its standard form, i.e.
\begin{align}\label{sbc2}
    T_{R}(\Tilde{u})-\bar{T}_{R}(\tilde{v})=T_{L}(\Tilde{u})-\bar{T}_{L}(\tilde{v}) \; \text{at} \; \Tilde{v}=\Tilde{u}
\end{align}
As noted earlier, (\ref{sbc2}) is an energy-conserving gluing condition. In other words, in the static frame, `$\Tilde{x}\Tilde{t}$' component of the stress tensor is continuous along the interface at $\Tilde{x}=0$. Even though it is not in the moving frame.
\\\\
\textit{\underline{Transmission and reflection coefficients}}: We will now compute the transmission ($\mathcal{T}$) and reflection ($\mathcal{R}$) coefficients for this dynamical interface.
 $\mathcal{T}$ and $\mathcal{R}$ are defined as the expectation values of ANEC operators with respect to the asymptotic states defined in section \ref{Collider Experiment}. We start with the transmission coefficient.
 In this setting, we will use conformal transformations (\ref{map}) to map the dynamical interface to a static interface and use (\ref{sbc2}).
\begin{equation}
\begin{split}
\mathcal{T}_{L } 
&=\lim_{D\to\infty}\frac{\int_{-\infty}^{\infty}du\braket{O_{L}^{1}(u_{1},D)|T_{R}(u)|O_{L}^{2}(u_{2},D)}_{I}}{\int_{-\infty}^{\infty}du\braket{O_{L}^{1}(u_{1},D)|T_{L}(u)|O_{L}^{2}(u_{2},D)}}
\end{split}
\end{equation}

The subscript $L$ in $\mathcal{T}$ denotes the origin of the incoming excitation. 
Instead of considering any generic primary or quasi-primary excitation, we will consider only chiral or anti-chiral operators here. The main reason for working with only the chiral or anti-chiral operators is that the general form of the three-point function of the stress tensor with two primary operators cannot be determined by symmetries alone, except when the operators are purely (anti) chiral\footnote{We need the exact form of the three-point function to evaluate the integrals explicitly.}. 

When defining our observables, we put the $D$ dependency explicitly. For the calculations, for the $CFT_{L}$, we will put the $D$ dependency on the $v$ coordinate, and for $CFT_{R}$, we will put the $D$ dependency on $u$ coordinate\footnote{Since, to reach the past infinity from the left side of the interface we need to take $v \rightarrow -\infty$ and to reach from the right side of the interface we need to take $u \rightarrow -\infty$}. In other words, if we define the excitation in $CFT_{R}$, the explicit  $D$ dependence will appear in $u$, not in $v$. In this way, we can take the $D \rightarrow \infty$ limit safely because all our calculations are done in the presence of either chiral or anti-chiral primaries. Hence, for the computations of $\mathcal{T}_{L}$ and $\mathcal{R}_{L}$ we may drop the explicit $D$ dependency, but, to emphasize its importance, we will keep it till the end of the calculations.\\ \\
{\textit{\underline{The numerator}:}
Let's look at the numerator after a generic conformal transformation  
\begin{equation}
\begin{split}
\braket{O_{L},D|\mathcal{E}_{R}|O_{_{L}},D}_{I}=& -\frac{1}{2 \pi} \int \left(\frac{du_{1}}{d\Tilde{u}_{1}}\right) d\Tilde{u}_{1} \Big{\langle} 0 \Big{|}f(p^{-1}(\Tilde{u}_{1})) \left(\frac{\partial\Tilde{u}_{1}}{\partial u_{1}}\right)^{h_{1}}_{u_{1}=p^{-1}(\Tilde{u}_{1})} O_{L}^{1}(\Tilde{u}_{1})\Big{|}\\\ & \int \left(\frac{du}{d\Tilde{u}}\right) d\Tilde{u} \left[ \left(\frac{\partial\Tilde{u}}{\partial u}\right)^{2}_{u=p^{-1}(\Tilde{u})} T_{R}(\Tilde{u})+\frac{c_{R}}{12} \{\Tilde{u},u\}\right]   \\& \Big{|}\int \left(\frac{du_{2}}{d\Tilde{u}_{2}}\right) d\Tilde{u}_{2} f(p^{-1}(\Tilde{u}_{2})) \left(\frac{\partial\Tilde{u}_{2}}{\partial u_{2}}\right)^{h_{2}}_{u_{2}=p^{-1}(\Tilde{u}_{2})} O^{2}_{L}(\Tilde{u}_{2})\Big{|}0\Big{\rangle}_{I}
\end{split}
\end{equation}
Subtracting the zero point energy, which is coming from the Schwarzian  part, we get 
\begin{equation}
\begin{split}
\braket{O_{L},D|\mathcal{E}_{R}|O_{_{L}},D}_{I}=
&-\frac{1}{2 \pi} \int \left(\frac{du_{1}}{d\Tilde{u}_{1}}\right) d\Tilde{u}_{1} \Big{\langle}0\Big{|}f(p^{-1}(\Tilde{u}_{1})) \left(\frac{\partial\Tilde{u}_{1}}{\partial u_{1}}\right)^{h_{1}}_{u_{1}=p^{-1}(\Tilde{u}_{1})} O_{L}^{1}(\Tilde{u}_{1})\\
&\Big{|}\int \left(\frac{du}{d\Tilde{u}}\right) d\Tilde{u}  \left(\frac{\partial\Tilde{u}}{\partial u}\right)^{2}_{u=p^{-1}(\Tilde{u})} T_{R}(\Tilde{u})\Big{|}\int \left(\frac{du_{2}}{d\Tilde{u}_{2}}\right) d\Tilde{u}_{2} f(p^{-1}(\Tilde{u}_{2})) \left(\frac{\partial\Tilde{u}_{2}}{\partial u_{2}}\right)^{h_{2}}_{u_{2}=p^{-1}(\Tilde{u}_{2})} O^{2}_{L}(\Tilde{u}_{2})\Big{|}0\Big{\rangle}_{I}
\end{split}
\end{equation}
To proceed further, we use the following identity: 
\begin{equation}\label{eq:identity}
\begin{split}
\Big{<}O_{L}^{1}(\Tilde{u}_{1})T_{R}(\Tilde{u})O_{L}^{2}(\Tilde{u}_{2})\Big{>}_{I}= \Bigg(\frac{c_{LR}}{c_{L}}\Bigg) \Big{<}O_{L}^{1}(\Tilde{u}_{1})T_{L}(\Tilde{u})O_{L}^{2}(\Tilde{u}_{2})\Big{>}
\end{split}
\end{equation}

This identity comes from the definition of OPE, which is valid even when a dynamical interface is present\footnote{we also don't put the $D$ dependency as it is true in general and doesn't depend on how one creates a state.}. The above identity can be derived for any general three-point correlation, e.g., where the fields are not necessarily (anti)chiral. For more details, we refer the readers to the reference \cite{Papadopoulos:2023kyd}. We will also need the form of three-point function of the stress tensor with two primary operators.  In the absence of an interface, this is given by:
\begin{equation}\label{eq:3ptfn}
\begin{split}
\Big{<}O_{L}^{1}(\Tilde{u}_{1})T_{L}(\Tilde{u})O_{L}^{2}(\Tilde{u}_{2})\Big{>}= h\Bigg{[}
&\frac{1}{(\Tilde{u}-\Tilde{u}_{1})^{2}(\Tilde{u}_{1}-\Tilde{u}_{2})^{2h}}+\frac{1}{(\Tilde{u}-\Tilde{u}_{2})^{2}(\Tilde{u}_{1}-\Tilde{u}_{2})^{2h}}\\
&-\frac{2}{(\Tilde{u}-\Tilde{u}_{1})(\Tilde{u}_{1}-\Tilde{u}_{2})^{2h+1}}-\frac{2}{(\Tilde{u}-\Tilde{u}_{2})(\Tilde{u}_{1}-\Tilde{u}_{2})^{2h+1}}\Bigg{]}
\end{split}
\end{equation}
It is zero unless $h_{1}=h_{2}=h$.  Finally, using \eqref{eq:3ptfn} and \eqref{eq:identity} we get
\begin{equation}
\begin{split}
\braket{O_{L},D|\mathcal{E}_{R}|O_{_{L}},D}_{I}&= -\frac{h}{2 \pi} \frac{c_{LR}}{c_{L}}\int \left(\frac{du_{1}}{d\Tilde{u}_{1}}\right) d\Tilde{u}_{1} f(p^{-1}(\Tilde{u}_{1})) \left(\frac{\partial\Tilde{u}_{1}}{\partial u_{1}}\right)^{h_{1}}_{u_{1}=p^{-1}(\Tilde{u}_{1})}\\
&\int \left(\frac{du_{2}}{d\Tilde{u}_{2}}\right) d\Tilde{u}_{2} f(p^{-1}(\Tilde{u}_{2})) \left(\frac{\partial\Tilde{u}_{2}}{\partial u_{2}}\right)^{h_{2}}_{u_{2}=p^{-1}(\Tilde{u}_{2})}\int \left(\frac{du}{d\Tilde{u}}\right) d\Tilde{u}  \left(\frac{\partial\Tilde{u}}{\partial u}\right)^{2}_{u=p^{-1}(\Tilde{u})}  \\
&\times\Bigg{[}\frac{1}{(\Tilde{u}-\Tilde{u}_{1})^{2}(\Tilde{u}_{1}-\Tilde{u}_{2})^{2h}}+\frac{1}{(\Tilde{u}-\Tilde{u}_{2})^{2}(\Tilde{u}_{1}-\Tilde{u}_{2})^{2h}}\\
&~~~~-\frac{2}{(\Tilde{u}-\Tilde{u}_{1})(\Tilde{u}_{1}-\Tilde{u}_{2})^{2h+1}}-\frac{2}{(\Tilde{u}-\Tilde{u}_{2})(\Tilde{u}_{1}-\Tilde{u}_{2})^{2h+1}}\Bigg{]}\,.
\end{split}
\end{equation}

Till now, all the discussions are valid for a generic interface. In what follows, we will restrict our discussions to the {\it escaping mirror profile} \eqref{escaping profile} 
This can be mapped to a static interface $\tilde{u}=\tilde{v}$ by the following conformal transformation
\begin{equation}
\begin{split}
\tilde{u}&=-\beta \log(1+e^{-u/\beta})\,,\qquad\tilde{v}=v\\
\Rightarrow u&=-\beta \log\left(e^{-\frac{\Tilde{u}}{\beta}}-1\right) \\
\frac{d\Tilde{u}}{du}&= \frac{e^{-u/\beta}}{1+e^{-u/\beta}} = 1-e^{\Tilde{u}/\beta},\qquad \frac{du}{d\Tilde{u}}= \frac{1}{1-e^{\Tilde{u}/\beta}}\,.
\end{split}
\end{equation}
If we evaluate the above integration, we will get \footnote{More details of the calculations in this section are given in appendix \ref{A1}}
\begin{equation}
\begin{split}
&\braket{O_{L},D|\mathcal{E}_{R}|O_{_{L}},D}_{I}\\
&=\lim_{\epsilon\to 0}\int d\Tilde{u}_{1}d\Tilde{u}_{2} f(p^{-1}(\Tilde{u}_{1})) f(p^{-1}(\Tilde{u}_{2})) \Bigg(\frac{\partial \Tilde{u}_{1}}{\partial u_{1}}\Bigg)^{h_{1}-1} \Bigg(\frac{\partial \Tilde{u}_{2}}{\partial u_{2}}\Bigg)^{h_{2}-1} \frac{2h}{(\Tilde{u}_{1}-\Tilde{u}_{2})^{2h}} \Bigg[ 4-2e^{\Tilde{u}_{1}/\beta}-2e^{\Tilde{u}_{2}/\beta}\Bigg] \frac{1}{\epsilon}
\end{split}
\end{equation}
We obtained the expression of the numerator without the $u_{1}$ and $u_{2}$ integrations. There is an $\epsilon$ in the above expression. We have used it to regulate the integration. Ultimately, we will take $\epsilon \rightarrow 0 $ limit. We will see that there will be an $\epsilon$  in the denominator also. These two $\epsilon$'s will cancel against each other, leaving us with a finite result. \\\\
\textit{\underline{The denominator}:}
We can follow the same steps for the denominator. However, the denominator is defined by the correlation function evaluated when no interface exists. So, we don't need to do the coordinate transformations. The expression of the denominator after doing the $u$-integration is given by
\begin{equation}
\begin{split}
\braket{O_{L},D|\mathcal{E}_{L}|O_{_{L}},D}= \lim_{\epsilon\to 0} -\frac{1}{2 \pi} \int du_{1} du_{2} f(u_{1}) f(u_{2}) \Bigg[\frac{2}{\epsilon} \frac{h}{(u_{1}-u_{2})^{2h}}\Bigg]
\end{split}
\end{equation}
Before going further, we need to choose some profiles of the wave packets. We choose the delta function as our wave packet
\begin{equation}
\begin{split}
f(u)=\delta(u-\ell)\,.
\end{split}
\end{equation}

$\ell$ is the point where the delta function is located, and our results crucially depend on $\ell$. Evaluating the integrations with the delta functions, we finally get the transmission coefficient for $CFT_{L}$
\begin{equation}\label{main1}
\begin{split}
\mathcal{T}_{L }=\frac{c_{LR}}{c_{L}}\Bigg( \frac{2}{1+e^{\frac{\ell}{\beta}}} \Bigg)
\end{split}
\end{equation}
We will now move on to the computation of the reflection coefficient
\begin{equation}
\begin{split}
\mathcal{R}_{L }&=\lim_{D\to\infty}\frac{\int_{-\infty}^{\infty}dv\braket{O_{L}^{1}(u_{1},D)|\Bar{T}_{L}(v)|O_{L}^{2}(u_{2},D)}_{I}-\int_{-\infty}^{\infty}dv\braket{O_{L}^{1}(u_{1},D)|\Bar{T}_{L}(v)|O_{L}^{2}(u_{2},D)}}{\int_{-\infty}^{\infty}du\braket{O_{L}^{1}(u_{1},D)|T_{L}(u)|O_{L}^{2}(u_{2},D)}} 
\end{split}
\end{equation}
As we have transformed only the $u$ coordinate, not the $v$ coordinate, the $\Bar{T}(v)$, which is the anti-chiral part of the stress tensor, will remain the same. But the operators, which are used to create the state, are chiral and will transform.  So, after the transformation in the numerator, we get 
\begin{equation}
\begin{split}
\braket{O_{L},D|\Bar{\mathcal{E}}_{L}|O_{_{L}},D}_{I}=& -\frac{1}{2 \pi} \int_{-\infty}^{0} \left(\frac{du_{1}}{d\Tilde{u}_{1}}\right) d\Tilde{u}_{1} \Big{\langle}0\Big{|}f(p^{-1}(\Tilde{u}_{1})) \left(\frac{\partial\Tilde{u}_{1}}{\partial u_{1}}\right)^{h_{1}}_{u_{1}=p^{-1}(\Tilde{u}_{1})} O_{L}^{1}(\Tilde{u}_{1})\Big{|} \int_{-\infty}^{\infty}  d\Tilde{v}  \hspace{.3cm} \Bar{T}_{L}(\Tilde{v}) \\
& \Big{|}\int_{-\infty}^{0} \left(\frac{du_{2}}{d\Tilde{u}_{2}}\right) d\Tilde{u}_{2} f(p^{-1}(\Tilde{u}_{2})) \left(\frac{\partial\Tilde{u}_{2}}{\partial u_{2}}\right)^{h_{2}}_{u_{2}=p^{-1}(\Tilde{u}_{2})} O^{2}_{L}(\Tilde{u}_{2})\Big{|}0\Big{\rangle}_{I}
\end{split}
\end{equation}
Though the numerator is different, the denominator will remain the same. The identity, which comes from the OPE, is 
\begin{equation}\label{eq:id2}
\begin{split}
\Big{<}O_{L}^{1}(\Tilde{u}_{1})\Bar{T}_{L}(\Tilde{v})O_{L}^{2}(\Tilde{u}_{2})\Big{>}_{I}= \Bigg(\frac{c_{L}-c_{LR}}{c_{L}}\Bigg) \Big{<}O_{L}^{1}(\Tilde{u}_{1})T_{L}(\Tilde{v})O_{L}^{2}(\Tilde{u}_{2})\Big{>} 
\end{split}
\end{equation}
%
Using \eqref{eq:id2}, we can do the integration, and the result is the following
\begin{equation}\label{main2}
\begin{split}
\mathcal{R}_{L } &= \Bigg(\frac{c_{L}-c_{LR}}{c_{L}}\Bigg)
\end{split}
\end{equation}
We are getting the same result as if the interface is static.


\eqref{main1} and \eqref{main2} are the main results of our paper. 
As we have seen earlier, for a static interface, the coefficients are fixed by the central charges of the two CFTs and the coefficient of the two-point functions between the left and the right stress tensors. They are not dependent on the specific choices of the state. But, this is not true for a dynamical interface.
In our dynamical situation, the transmission and reflection coefficients are given by 
\begin{equation}
\begin{split}
\mathcal{T}_{L }=\frac{c_{LR}}{c_{L}}\Bigg( \frac{2}{1+e^{\frac{\ell}{\beta}}} \Bigg)\,, \quad
\mathcal{R}_{L }= \frac{c_{L}-c_{LR}}{c_{L}}\,,\quad
\Rightarrow\,\,\mathcal{T}_{L}+\mathcal{R}_{L }= 1- \frac{c_{LR}}{c_{L}}\Bigg( \frac{e^{\frac{\ell}{\beta}}-1}{e^{\frac{\ell}{\beta}}+1} \Bigg)
\end{split}
\end{equation}
A few comments are in order:
\begin{itemize}
    \item The transmission coefficient changes due to the dynamical interface, but the reflection coefficient remains the same as the static interface. 
    \item The transmission coefficient explicitly depends on $\beta$, but the reflection coefficient does not. In the limit ($\beta \rightarrow 0$), there is only reflection, no transmissions.
    \item The reflection and transmission coefficients are always less than one, which respects unitarity \cite{Billo:2016cpy}. If we add the two coefficients, it is one only when we take the ($\beta \rightarrow \infty$) limit.  
\end{itemize}
%
%
%
%
%
%
%
We can similarly compute $\mathcal{T}_R$ and $\mathcal{R}_R$. Here, we will quote the result (see appendix (\ref{A1}) for details).
\begin{equation}
\begin{split}
\mathcal{T}_{R } = \frac{c_{LR}}{c_{R}}, \hspace{.1cm} \mathcal{R}_{R } =  \frac{c_{R}-c_{LR}}{c_{R}} \Bigg( \frac{2}{1+e^{\frac{\ell}{\beta}}} \Bigg) \hspace{.1cm} \Longrightarrow \mathcal{T}_{R}+\mathcal{R}_{R}= \frac{2}{1+e^{\frac{\ell}{\beta}}} +\frac{c_{LR}}{c_{R}}\Bigg( \frac{e^{\frac{\ell}{\beta}}-1}{e^{\frac{\ell}{\beta}}+1} \Bigg)
\end{split}
\end{equation}
When the origin of the incoming excitation is in the right CFT ($CFT_R$), we find:
\begin{itemize}
        \item The reflection coefficient changes due to the dynamical interface, but the transmission coefficient remains the same. We saw the opposite for the $CFT_{L}$.
         \item Now, the reflection coefficient explicitly depends on $\beta$. In ($\beta \rightarrow 0$) limit, there is only transmissions, no reflection (again, this is the opposite of what happens for $CFT_{L}$). 
        \item The reflection and transmission coefficients are always less than one, as expected from unitarity. if we add the two coefficients, it is one only when we take ($\beta \rightarrow \infty$).  
       \item In the ($\beta \rightarrow 0$) limit, there is only (partial) reflection, no transmissions from the left CFT, but for the right CFT, there is only (partial) transmission, no reflection. This means the moving mirror behaves as a semipermeable membrane as $\beta\rightarrow 0$. However, it should be noted that in neither case do we get pure transmission or reflection since the transmission and reflection coefficients are always less than one, which implies that the interface absorbs energy.  
\end{itemize}
%
%
%
%
%
Both $\mathcal{T}_L+\mathcal{R}_L$ and $\mathcal{T}_R+\mathcal{R}_R$ are less than one, which implies energy is getting lost due to absorption by the interface. If we add $\mathcal{T}_L$, $\mathcal{R}_L$, $\mathcal{T}_R$ and $\mathcal{R}_R$ we get
\begin{equation}
\begin{split}
\mathcal{T}_{L}+\mathcal{R}_{L }+\mathcal{T}_{R}+\mathcal{R}_{R}= 2-\Bigg( \frac{e^{\frac{\ell}{\beta}}-1}{e^{\frac{\ell}{\beta}}+1} \Bigg)\Bigg(1+\frac{c_{LR}}{c_{L}}-\frac{c_{LR}}{c_{R}}\Bigg)
\end{split}
\end{equation}
As, $c_{LR}/c_L-c_{LR}/c_R$ always lies within $-1$ and $+1$ the above quantity is always less than two. This quantifies the amount of energy absorbed at the interface when we throw excitations from both sides toward the interface at the same time.
%
%
We end this section with some comments on some special solutions to the gluing condition.
\begin{itemize}


    \item For factorizing gluing condition in the case of the static interface, one must have $c_{LR} = 0$. Then, for the moving interface, we get, $\mathcal{T}_{L} =0$, $\mathcal{R}_{L}=1$. and $\mathcal{T}_{R} =0$ and $\mathcal{R}_{R}=\frac{2}{1+e^{{\ell}/{\beta}}}$. In $\beta \rightarrow 0$, $\mathcal{R}_{R} \rightarrow 0$. Hence, for factorizing cases, the interface behaves like a one-sided mirror. The right side is completely frozen in $\beta \rightarrow 0$.

    \item For transparent static interfaces, one has $c_{LR}=c_{L}=c_{R}$. Substituting this value in our final result, we end up with one-sided transmission, i.e., $\mathcal{T}_R =1$ and $\mathcal{R}_R=\mathcal{T}_L =\mathcal{R}_L =0$\footnote{Naively, one might expect that transparent boundary condition should be akin to having no boundary at all, and so one should expect completely transparent transmission from either side. However, in this case, the non-trivial conformal transformation to the escaping profile creates a geometry in which one gets one-sided transmission across the topological interface.}.



\end{itemize}



\subsection{ A different gluing condition}\label{diffbdy}
In earlier sections, we have mostly engaged with the natural ICFT boundary condition of the form \eqref{sbc1}
In this section, however, we discuss the same scattering problem with a different set of interface gluing conditions. 
\begin{align}\label{bc1}
T_{R}(\tilde{u})+T_{L}(\tilde{u}) = \bar{T}_{R}(\tilde{v})+\bar{T}_{L}(\tilde{v}), \; \text{at} \; \tilde{v}=\tilde{u}\,.
\end{align}

 We could get (\ref{bc1}) from the ICFT gluing condition (\ref{standard icft}), by interchanging either $T_{L}(u) \leftrightarrow \bar{T}_{L}(v)$ or by $T_{R}(u)\leftrightarrow \bar{T}_{R}(v)$, and so one can interpret this as a gluing condition at the interface between $CFT_L$ and $\overline{CFT}_R$. Equivalently, in the folded picture, this may also be naturally thought of as the conformal boundary condition imposed at the boundary of a tensor product BCFT of $CFT_L\otimes CFT_R$. 

As before, we get to the case of the moving mirror profile by doing a conformal transformation, after which the boundary condition corresponding to (\ref{bc1}) will be of the form\footnote{Here again, we subtracted the vacuum energy part of the stress tensor. In other words here $T \rightarrow T-\langle T \rangle$.}
\begin{align}\label{bc4}
(p'(u))^{-2}(T_{R}(u)+T_{L}(u))=\bar{T}_{R}(v)+\bar{T}_{L}(v) \; \text{at} \; v=p(u)
\end{align}
 From (\ref{bc4}), it is again clear that the boundary condition in the moving frame is not conformal.  \\\\
\textit{\underline{Transmission and Reflection Coefficients}:}
 For (\ref{bc1}) or (\ref{bc4}), we will compute transmission and reflection coefficients. As mentioned earlier, the gluing condition is most naturally understood as a conformal boundary condition for a tensor product  BCFT - CFTs $CFT_{L}\otimes CFT_{R}$, both of which live on the same side of the boundary. For instance, we can choose both $CFT_{L,R}$ live on the Left(L) side$(\Tilde{x} <0)$. In this setup, we can define transmission coefficients(which determine the transmission of energy from one CFT to the other) as follows:
 \begin{align}
   \mathcal{T}_{L} = \lim_{D\rightarrow \infty} \frac{\langle \mathcal{O}_{L,D}| \int^{\infty}_{-\infty}dv\,\bar{T}_{R}(v)|\mathcal{O}_{L,D}\rangle_{I}}{\langle \mathcal{O}_{L,D}| \int^{\infty}_{-\infty}du\,T_{L}(u)|\mathcal{O}_{L,D}\rangle} ,\quad \; \mathcal{T}_{R} = \lim_{D\rightarrow \infty} \frac{\langle \mathcal{O}_{R,D}| \int^{\infty}_{-\infty}dv\,\bar{T}_{L}(v)|\mathcal{O}_{R,D}\rangle_{I}}{\langle \mathcal{O}_{R,D}| \int^{\infty}_{-\infty}du\,T_{R}(u)|\mathcal{O}_{R,D}\rangle}\,.  
 \end{align}
 The reflection coefficients will be: $\mathcal{R}_{L} = 1-\frac{c_{L\bar{R}}}{c_{L}}, \mathcal{R}_{R}=1-\frac{c_{R\bar{L}}}{c_{R}}$. Here we defined $\langle T_{L}\bar{T}_{R}\rangle \propto c_{L\bar{R}}$ and $\langle \bar{T}_{L}T_{R}\rangle \propto c_{\bar{L}R}$\footnote{However, in this gluing condition $\langle T_{L} T_{R} \rangle = 0$ which can be shown by exchanging $T_{R} \leftrightarrow \bar{T}_{R}$ in the usual ICFT gluing (\ref{standard icft}).}. For the special subset of gluing condition:
 \begin{align}\label{bc3}
T_{L} = \bar{T}_{R}, \; \bar{T}_{L}=T_{R} \;  \text{at} \; \tilde{u}=\tilde{v}
\end{align}
We can immediately see $c_{L\bar{R}} = c_{L}=\bar{c}_{R}$ and $c_{\bar{L}R}=\bar{c}_{L}=c_{R}$. This gluing condition is studied in great detail in Appendix \ref{app:B}. Hence for this case, $\mathcal{T}_{L,R} = 1$ and $\mathcal{R}_{L,R}=0$. In other words, all the energy from one CFT will be transmitted to the other one in this gluing condition. \\

We could also interpret (\ref{bc1}) as a gluing condition at the interface of  $ CFT_L$ and $\overline{CFT}_R$ living on either side of the interface. The interchange of $CFT_R$ with $\overline{CFT}_R$ suggests that we set up a slightly modified scattering experiment as follows: When we create an excitation in the Left(L) region($\tilde{x}<0$) and scatter it to the Right(R) region($\tilde{x}>0$), in $CFT_{R}$ we interchange $T_{R} \leftrightarrow \bar{T}_{R}$. Similarly, when we send an excitation from R to L, we must interchange $T_{L} \leftrightarrow \bar{T}_{L}$ \footnote{ At this stage, we have no precise understanding of what such a chiral-anti chiral exchange implies physically.}. Mathematically, this amounts to modifying the definitions of the transmission coefficients. %
%
\begin{align}\label{modified transport}
\mathcal{\widetilde{T}}_{L} = \lim_{D\rightarrow \infty} \frac{\langle \mathcal{O}_{L,D}| \int^{\infty}_{-\infty}dv\,\bar{T}_{R}(v)|\mathcal{O}_{L,D}\rangle_{I}}{\langle \mathcal{O}_{L,D}| \int^{\infty}_{-\infty}du\,T_{L}(u)|\mathcal{O}_{L,D}\rangle} ,\quad \; \mathcal{\widetilde{T}}_{R} = \lim_{D\rightarrow \infty} \frac{\langle \mathcal{\bar{O}}_{R,D}| \int^{\infty}_{-\infty}du\,T_{L}(u)|\mathcal{\bar{O}}_{R,D}\rangle_{I}}{\langle \mathcal{\bar{O}}_{R,D}| \int^{\infty}_{-\infty}dv\, \bar{T}_{R}(v)|\mathcal{\bar{O}}_{R,D}\rangle}\,.
\end{align}
The reflection coefficients will remain unchanged. 
Using a similar computation as before \footnote{Since the computation is analogous to the previous case, we are not presenting here.}, we get
\begin{align}\label{coeff bc1}
\mathcal{\widetilde{T}}_{L} = \frac{c_{L\bar{R}}}{c_{L}}, \; \mathcal{R}_{L}=1-\frac{c_{L\bar{R}}}{c_{L}}, \;\; \mathcal{\widetilde{T}}_{R}= \frac{c_{L\bar{R}}}{\bar{c}_{R}}\left(\frac{2}{1+e^{\frac{\ell}{\beta}}}\right), \; \mathcal{R}_{R} = \left(1-\frac{c_{L\bar{R}}}{\bar{c}_{R}}\right)\frac{2}{1+e^{\frac{\ell}{\beta}}}\,.
\end{align}
 \\
\textit{\underline{$\beta \rightarrow 0$ limit and a horizon connection}:}
In the $\beta \rightarrow 0$ limit, the above coefficients (\ref{coeff bc1}) reduces to
\begin{align}
\mathcal{\widetilde{T}}_{L} = \frac{c_{L\bar{R}}}{c_{L}}, \; \mathcal{R}_{L}=1-\frac{c_{L\bar{R}}}{c_{L}}, \;\; \mathcal{\widetilde{T}}_{R} \rightarrow 0, \; \mathcal{R}_{R} \rightarrow 0\,.
\end{align} 
Hence, from the Right side, nothing can transmit and reflect in this limit. Now let us comment on some features in the $\beta \rightarrow 0$ limit.
\begin{itemize}
\item If we consider the special case of total transmission (\ref{bc3}), we can see $\mathcal{\widetilde{T}}_{L}=1$ and $\mathcal{R}_{L}=\mathcal{R}_{R}=0$. Also in the $\beta \rightarrow 0$, $\mathcal{\widetilde{T}}_{R} \rightarrow 0$. Hence, this boundary condition makes the interface to be one-sided transparent subjected to the scattering experiment defined as (\ref{modified transport}). 
  
\item 
Following \cite{Akal:2021foz}, one can also compute the Bogoliubov coefficients between in and out modes constructed on past and future null infinities, respectively. In this setting, if we consider copies of scalar fields subjected to (\ref{bc3}), the computation of an average number of particle production at future null infinity will be exactly similar to that in \cite{Akal:2021foz,Birrell:1982ix}. Thus, using (\ref{bc3}) one can reproduce Hawking radiation in $\beta\rightarrow 0$ limit. 

 \item Also as mentioned earlier, in $\beta \rightarrow 0$, the moving interface $v=p(u)$ approaches to null line $v=0$. 
  \end{itemize}
  The above three facts suggest that at $\beta \rightarrow 0$, the interface subjected to (\ref{bc3}), behaves like a horizon. In the next section, we will connect this result with the eternal black hole.
\section{Connection to eternal black hole}\label{sec:Horizon}
The analysis of the previous section suggests an analogy of the {\it moving interface} with the eternal black hole or wormhole geometry. Under the gluing condition (\ref{bc3}), we have observed that the L region behaves like a left exterior, and the R region behaves like an interior of a black hole subjected to a chiral anti-chiral exchange in the $CFT_{R}$. We can similarly model an interior with respect to the right exterior using the following boundary conditions:
\begin{align}\label{bc3-II}
T_{L}(\tilde{u}')=\bar{T}_{R}(\tilde{v}'); \; T_{R}(\tilde{u}') = \bar{T}_{L}(\tilde{v}'), \; \text{at}\; \tilde{v}'=\tilde{u}';\; \text{where}\; \tilde{v}'=p(v) \; \text{and} \; \tilde{u}'=u
\end{align}
This means if we choose the interface profile as $u=p(v)$, then (\ref{bc3-II}) implies one-sided transmission from R to L side, in the $\beta\rightarrow 0$ limit, following the same scattering experiment (\ref{modified transport}) as described in the last section. However, to mimic a connecting wormhole-like geometry, we need to provide evidence that it has properties similar to that of a wormhole geometry. For example, we must understand Hawking radiation in future infinities of the exterior part. In this section, we would like to make some connections regarding this. Before proceeding, let us briefly point out some key features of eternal two-sided black holes.\\\\
\textit{\underline{TFD and eternal black hole}:}
An eternal two-sided black hole or a black hole in thermal equilibrium is often characterized by the existence of the thermal vacuum, which is the so-called Hartle-Hawking-Israel state \cite{PhysRevD.28.2960,Israel:1976ur}. This is a maximally entangled state constructed out of the eigenstates associated with the fock space of two exterior regions of the Penrose diagram. Most notably in AdS/CFT \cite{Maldacena:1997re}, this is described by a thermofield double (TFD) state constructed out of two copies of CFTs\cite{Maldacena:2001kr}. The entangled structure of the state is solely responsible for thermal Hawking radiation detected by asymptotic observers of each region. Geometrically, the non-zero connected correlators in such a TFD state are an artifact of the connectedness of the spacetime through a smooth interior. In modern terminology, this feature is described as ER=EPR \cite{Maldacena:2013xja}, where the wormhole or Einstein-Rosen bridge is the geometrization of a maximally entangled TFD state. An independent construction of TFD state or modeling a wormhole solution in order to study the validity of ER=EPR as well as to understand the dynamics of the black hole interior is an area of active research. Some key properties of TFD state are as follows  \cite{Hartman}.
\begin{itemize}
\item The maximally entangled TFD state in a Hilbert space $\mathcal{H}=\mathcal{H}_{R}\otimes \mathcal{H}_{L}$ can be written as 
\begin{align}
|\psi_{TFD}\rangle = \frac{1}{\sqrt{Z(\beta)}}\sum_{n}e^{-\beta E_{n}}|n_{L}\rangle |n_{R}\rangle
\end{align}
Where $|n_{L,R}\rangle$ are eigenstates of Hamiltonian $H_{L,R}$.
\item Two-sided connected correlators are non-vanishing. 
\begin{align}
\langle\mathcal{O}_{L}\mathcal{O}_{R}\rangle_{TFD}-\langle\mathcal{O}_{L}\rangle_{TFD}\langle\mathcal{O}_{R}\rangle_{TFD} \neq 0
\end{align}
\item $|\psi_{TFD}\rangle$ remains time independent for evolution under $H_{R}-H_{L}$. However, if we evolve it with $H_{R}+H_{L}$, then the state becomes time-dependent.
\begin{align}
|\psi(t)\rangle_{TFD} = \sum_{n}e^{-\beta E_{n}-2iE_{n}t}|n_{L}\rangle |n_{R}\rangle
\end{align}
The dynamics of the black hole can be studied in such a time-dependent way (interesting physics might happen after much larger than scrambling time. In fact, most of the interior dynamics could be captured by the late time limit of the time-dependent TFD state.
\item The consequence of the above fact is geometrized as the time direction of the eternal black hole in the right side is upward while that on the left side is downward. Left-Right correlator can be rewritten as Left-Left correlator with certain analytic continuation in $t \rightarrow t-i\beta$
\begin{align}
\langle\mathcal{O}_{L}(x_{1},0)\mathcal{O}_{L}(x_{2},t)\rangle_{TFD} = \langle\mathcal{O}_{L}(x_{1},0)\mathcal{O}^{\dagger}_{R}(x_{2},t-i\beta)\rangle_{TFD}
\end{align}
\end{itemize} 
\textit{\underline{Bogoliubov coefficients}:}
%
Here, we want to make a connection between two exteriors of a black hole and $CFT_{R,L}$. We start with a tensor product of two copies of CFTs (L and R) satisfying the boundary condition (\ref{bc3}) at the static boundary. 
The mapping from moving ($u,v$) to static frame ($\tilde{u},\tilde{v}$) does not cover the entire spacetime in $(\tilde{u},\tilde{v})$ coordinate. Since $p(u) <0$ for any value of $u$, $\tilde{u}$ is always negative ($\tilde{u}<0$). The mapping tells that the boundary maps to $\tilde{x} = 0, \tilde{t} <0$. The right region (R) corresponds to ($\tilde{x}>0\, \cup\, \tilde{u}<0$) . Initially, the two CFTs in this description reside in R region. Now we consider a parity transformation ($\tilde{x}\rightarrow -\tilde{x}$)  in $CFT_{L}$ such that it maps to the left region (L) ($\tilde{x}<0\,  \cup\, \tilde{v}<0$). However, this transformation will interchange $T_{L}\leftrightarrow \bar{T}_{L}$. If we change the time direction in L to downwards, i.e., $(\tilde{t} \rightarrow -\tilde{t})$, then in L $\tilde{u} \rightarrow -\tilde{u}$ and $\tilde{v} \rightarrow -\tilde{v}$. This again keeps the boundary condition (\ref{bc3}) intact\footnote{One should keep in mind, here $T_{L}(\tilde{u}) \implies T_{L}(-\tilde{u})$. Also, this opposite time direction enforces continuity of $T_{\tilde{x}\tilde{t}}$ in both sides}. Note that this construction of  $CFT_{L}$ in L is reminiscent of $CRT$ transformation \cite{Haag:1992hx,Witten_2018} in the usual Rindler or two-sided black hole picture. However, the anti-unitary operator $J=CRT$ transforms an algebra of operators to its commutant \cite{Haag:1992hx,Witten_2018}. Usually, a commutant is causally inaccessible to the other side. Our construction is different from that, where the CFTs can interact via the interface. In $\beta \rightarrow 0$ limit, this transformation from R to L corresponds to the following analytic continuation from right to left in ($u,v$) coordinates: $u \rightarrow u+i\pi\beta$ and $v \rightarrow -v$. Note that if $v=0$, this exactly translates to $t \rightarrow t+i\pi\beta$.


We start by considering two copies of free massless scalar field theory denoted as $\phi_{L,R}$ in L and R region. Since we are considering free massless scalars, they can move freely in left-moving $\phi^{-}(v)$ and right-moving $\phi^{+}(u)$ component before reaching the interface. At the interface, they interact, maintaining the gluing condition (\ref{bc3}) in a static frame. One of those solutions is referred as the standard positive frequency ($\omega$) incoming modes $\phi^{in}_{(R,L),\omega}$ having both left moving and right moving sectors, defined on $\mathcal{I}^{-}_{R}$ and $\mathcal{I}^{+}_{L}$ and we have an expansion of fields in terms of these positive frequency incoming modes  as the following:
\begin{align}\label{phi in}
\phi(u,v) = \int^{\infty}_{0}d\omega \Big[a_{R,\omega}\phi^{in}_{R,\omega}+a_{R,\omega}^{\dagger}\phi^{in*}_{R,\omega} + a_{L,-\omega}\phi^{in}_{L,-\omega} + a_{L,-\omega}^{\dagger}\phi^{in*}_{L,-\omega}\Big]
\end{align}
where $a,a^{\dagger}$ are annihiliation and creation operators such that $a_{R,L}|0\rangle_{in} = 0$. Note that, at null infinities, the asymptotic structures of the fields are essentially the same in both moving and static frames, as the boundary effect is negligible. $|0\rangle_{in}$ is the vacuum defined on $\mathcal{I}^{-}_{R}$ and $\mathcal{I}^{+}_{L}$. The normal ordered stress tensor for free scalar has the form $T^{\pm} =:\partial_{\pm}\phi\partial_{\pm} \phi :$ \footnote{Here $+$ refers to $\tilde{u}$ and $-$ refers to $\tilde{v}$.}. Since scalar field equation is unchanged under 2$D$ conformal transformation, we can expand any mode solution $\phi_{\omega}(\tilde{u},\tilde{v})$ in terms of positive frequency modes $\phi^{+},\phi^{-}$ as $\phi_{\omega}=(\phi^{+}_{\omega}+\phi^{-}_{\omega})$ and $\phi = \int^{\infty}_{0}d\omega [a_{\omega}\phi_{\omega}+a_{\omega}^{\dagger}\phi^{*}_{\omega}]$, the solution of (\ref{bc3}) is simply of the form\footnote{Here we took the expectation value of the stress tensor in corresponding vacuum}:
\begin{align}\label{scalar bc 1}
&\int^{\infty}_{0}d\omega \partial_{+}\phi^{+}_{L,-\omega} \partial_{+}\phi^{+*}_{L,-\omega} = \int^{\infty}_{0}d\omega' \partial_{-}\phi^{-}_{R,\omega'} \partial_{-}\phi^{-*}_{R,\omega'} \; \text{at} \; \tilde{v}=\tilde{u}\\
& \int^{\infty}_{0}d\omega \partial_{+}\phi^{+}_{R,\omega} \partial_{+}\phi^{+*}_{R,\omega} = \int^{\infty}_{0}d\omega' \partial_{-}\phi^{-}_{L,-\omega'} \partial_{-}\phi^{-*}_{L,-\omega'} \; \text{at} \; \tilde{v}=\tilde{u} 
\end{align} 
We take one simple solution of the above boundary condition in the following form
\begin{align}\label{scalar bc2}
&\partial_{+}\phi^{+}_{L,-\omega} = \partial_{-}\phi^{-}_{R,\omega} \; \text{at} \; \tilde{v}=\tilde{u}\\
&\partial_{+}\phi^{+}_{R,\omega} = \partial_{-}\phi^{-}_{L,-\omega} \; \text{at} \; \tilde{v}=\tilde{u}
\end{align}
Since in left region, time is reversed or $\tilde{u} \rightarrow -\tilde{u}$ and $\tilde{v} \rightarrow -\tilde{v}$, the positive frequency mode in the left region should be understood as $\phi_{L,-\omega}$.\footnote{Also in general one could also have $\partial_{+}\phi^{+}_{L,\omega} = -\partial_{-}\phi^{-}_{R,\omega}$ and $\partial_{+}\phi^{+}_{R,\omega} = -\partial_{-}\phi^{-}_{L,\omega}$. However, we can check this does not affect the end result, and for the physical interface, the plus sign will be more natural, as we will see later.} The opposite direction of time in both sides maintains the continuity of the stress tensor across the interface. We will now construct the incoming and outgoing modes in R region.\\
 The incoming modes $\phi^{in}_{R}$ defined in $\mathcal{I}^{-}_{R}$ has both left going and right going modes. The standard left going modes must have the form $e^{-i\omega v}$ all over the region from $\mathcal{I}^{-}_{R}$(for $v<0$ region) upto the interface\cite{Birrell:1982ix}. The incoming mode from $v>0\,(\tilde{v}>0)$ region of $\mathcal{I}^{-}_{R}$ will never reach  the interface $v=p(u)\, (\tilde{v}=\tilde{u})$. These are called trapped modes and they end in $(\mathcal{I}^{+}_{L} \cap (v>0))$. We will neglect these modes in the rest of the note as those are not important in the connection to Hawking radiation \cite{PhysRevD.36.2327}. However, the right-moving modes will be modified due to the Doppler effect from the surface of the accelerating interface due to the boundary condition (\ref{scalar bc2})\footnote{More precisely, the counterpart of (\ref{scalar bc2}) in $(u,v)$ frame.}. In static frame, (\ref{scalar bc2}) implies the right moving part will be fixed as $\phi^{+} = e^{-i\omega \tilde{u}}$\cite{Davies:1976hi}. Thus in $(u,v)$ coordinate the form is fixed as $e^{-i\omega p(u)}$. Hence, we can uniquely fix the positive frequency incoming mode for both left and right regions as
\begin{align}\label{in}
\phi^{in}_{R,\omega} = \phi^{in}_{L,-\omega}= \frac{i}{\sqrt{4\pi\omega}}\Big(e^{-i\omega v}\theta(-v) + e^{-i\omega p(u)}\Big)
\end{align}
One can similarly find the outgoing $\phi^{out}$ modes in $\mathcal{I}^{+}_{R}$. The standard right moving modes will be of the form $e^{-i\omega u}$ from $\mathcal{I}^{+}_{R}$ to the interface if we trace back it in time. Hence in static frame $(\tilde{u},\tilde{v})$, the right moving mode is $e^{-i\omega p^{-1}(\tilde{u})}$. However, the left moving modes, which are sourced from the $\phi_{L}$, will be complicated again due to the boundary condition of accelerating interface trajectory. The positive frequency right moving part of outgoing mode in $\mathcal{I}^{-}_{L}$ is also having the same form $e^{-i\omega p^{-1}(\tilde{u})}$. Thus the boundary condition (\ref{scalar bc2}) fixes the left moving part as $\phi^{-} = e^{-i\omega p^{-1}(\tilde{v})}$. Thus in $(u,v)$ coordinate the form is fixed as $e^{-i\omega p^{-1}(v)}$. Thus, using boundary condition, we can uniquely fix the left moving sector of $\phi^{out}$, and we get the positive frequency out mode as
\begin{align}\label{out}
\phi^{out}_{R,\omega} = \phi^{out}_{L,-\omega} = \frac{i}{\sqrt{4\pi\omega}}\left(e^{-i\omega p^{-1}(v)}\theta(-v)+e^{-i\omega u}\right)
\end{align}
Note that the normalization of $\phi^{out}_{R,L}$ and $\phi^{in}_{R,L}$ are fixed using normalizability of Klein-Gordon inner product at Cauchy slices $\mathcal{I}^{+}$ and $\mathcal{I}^{-}$ which are of the form:
\begin{align}\label{KG}
&(\phi^{in}_{R,\omega},\phi^{in}_{R,\omega'})_{K.G} = -i\int^{\infty}_{-\infty} dv \left(\phi^{in}_{R,\omega}\partial_{v}(\phi^{in}_{R,\omega'})^{*}- (\phi^{in}_{R,\omega'})^{*}\partial_{v}\phi^{in}_{R,\omega}\right) = \delta(\omega-\omega') \\
&(\phi^{out}_{R,\omega},\phi^{out}_{R,\omega'})_{K.G} = -i\int^{\infty}_{-\infty} du \left(\phi^{out}_{R,\omega}\partial_{u}(\phi^{out}_{R,\omega'})^{*}- (\phi^{out}_{R,\omega'})^{*}\partial_{u}\phi^{out}_{R,\omega}\right) = \delta(\omega-\omega') 
\end{align}
 
The interesting fact is that the expression for $\phi^{in,out}_{R}$ are the same up to a relative negative sign between the left and right moving sectors, with that of one-sided moving mirror case with Dirichlet boundary condition at mirror\cite{Birrell:1982ix}. In the mirror case, we demand $\phi=0$ at the mirror location\footnote{Note that if we would like to construct in and out state in right region for the topological interface boundary condition, i.e., $T_{L}=T_{R}, \; \bar{T}_{L}=\bar{T}_{R} \; \text{at}\; \tilde{v}=\tilde{u}$, we need to satisfy the following boundary condition on the scalar fields: $\partial_{+}\phi^{+}_{R}=\partial_{+}\phi^{+}_{L}, \partial_{-}\phi^{-}_{R}=\partial_{-}\phi^{-}_{L}$. Thus, the left(right)moving part will always be Left(right) moving, and there will be no Doppler effect due to the nature of the interface. For this case, one can argue the positive frequency modes are $\phi^{R}_{in} = e^{-i\omega v}+e^{-i \omega u} = \phi^{R}_{out}$. Thus, there will be no particle creation for the topological interface.}. 

We do not need to impose this in our case since the modes of one copy are transmitted into the other one. That is why we do not use the solution of boundary condition $\partial_{+}\phi^{+}_{L,-\omega} = -\partial_{-}\phi^{-*}_{R,\omega}$ and $\partial_{+}\phi^{+}_{R,\omega} = -\partial_{-}\phi^{-*}_{L,-\omega}$. However, we will see that we end up getting similar Bogoliubov coefficients for this case, and the expectation value of the number operator precisely matches that for a single-sided black hole at a sufficiently late time.\\
Similar to (\ref{phi in}), the field $\phi$ can also be expanded in terms of outgoing modes respectively as 
\begin{align}\label{phi out}
\phi(u,v) = \int^{\infty}_{0}d\omega \left[b_{R,\omega}\phi^{out}_{R,\omega}+b_{R,\omega}^{\dagger}\phi^{out*}_{R,\omega} + b_{L,-\omega}\phi^{out}_{L,-\omega} + b_{L,-\omega}^{\dagger}\phi^{out*}_{L,-\omega}\right]
\end{align}
Here $b_{\omega}$'s are annihilation operators and $b^{\dagger}_{\omega}$'s are creation operators. Using Bogoliubov transformation, we can write positive frequency L(R) out modes in terms of positive frequency incoming modes of R(L). This decomposition is natural from the boundary condition imposed on the modes:
\begin{align}
\phi^{out}_{R,\omega} = \int^{\infty}_{0}d\omega'(\alpha^{RL*}_{\omega\omega'}\phi^{in}_{L,-\omega'}-\beta^{RL}_{\omega\omega'}\phi^{in*}_{L,-\omega'})
\end{align}
A similar expression also holds for the left(L) sector as
\begin{align}
\phi^{out}_{L,-\omega} = \int^{\infty}_{0}d\omega'(\alpha^{LR*}_{\omega\omega'}\phi^{in}_{R,\omega'}-\beta^{LR}_{\omega\omega'}\phi^{in*}_{R,\omega'})
\end{align}
We can evaluate the Bogoliubov coefficients $\alpha_{\omega\omega'},\beta_{\omega\omega'}$ as
\begin{align}
\alpha^{IJ}_{\omega\omega'} = (\phi^{out}_{I,\omega},\phi^{in}_{J,\omega'})_{K.G}\,, \quad \beta_{\omega\omega'}^{IJ}= -(\phi^{out}_{I,\omega},\phi^{in*}_{J,\omega'})_{K.G} \; \,\,\text{ for}\; (I,J)=R,L\,.
\end{align}
We can, in principle, invert the above relations, i.e., writing $\phi^{in}_{R,L}$ in terms of $\phi^{out}_{R,L}$ as the following:
\begin{align}\label{in out L1}
\phi^{in}_{R,\omega} = \int^{\infty}_{0}d\omega'(\alpha^{LR}_{\omega\omega'}\phi^{out}_{L,-\omega'}+\beta^{LR}_{\omega\omega'}\phi^{out*}_{L,-\omega'})
\end{align}
This is possible when the following relations are true\footnote{There might be a constant factor appearing in the RHS, which we can absorb and redefine Bogoliubov coefficients accordingly.}:
\begin{align}\label{bogo const 1}
 &\int^{\infty}_{0}d\omega' ( |\alpha^{LR}_{\omega\omega'}|^{2} - |\beta^{LR}_{\omega\omega'}|^{2}) = \delta(\omega-\omega')  
\end{align}
Similarly we could also express $\phi^{in}_{L}$ in terms of $\phi^{out}_{L,R}$ as the following:
\begin{align}\label{in out R1}
\phi^{in}_{L,-\omega} = \int^{\infty}_{0}d\omega'(\alpha^{RL}_{\omega\omega'}\phi^{out}_{R,\omega'}+\beta^{RL}_{\omega\omega'}\phi^{out*}_{R,\omega'})
\end{align}
This is also true given some constraints of Bogoliubov coefficients as in (\ref{bogo const 1}).
\begin{align}\label{bogo const 2}
 &\int^{\infty}_{0}d\omega' ( |\alpha^{RL}_{\omega\omega'}|^{2} - |\beta^{RL}_{\omega\omega'}|^{2}) = \delta(\omega-\omega')  
\end{align}

Using (\ref{in out L1}) and (\ref{in out R1}), one can see that the number operator expectation value, which is defined as the average number of particles created at $\mathcal{I}^{+}_{R}$ and $\mathcal{I}^{-}_{L}$ is related to $\beta^{RL}_{\omega\omega'}$ and $\beta^{LR}_{\omega\omega'}$ as
\begin{align}
&b_{R,\omega}|0_{in}\rangle = a_{L,-\omega}^{\dagger}\beta^{RL*}_{\omega\omega'}|0_{in}\rangle, \quad b_{L,-\omega}|0_{in}\rangle = a_{R,\omega}^{\dagger}\beta^{LR*}_{\omega\omega'}|0_{in}\rangle\,, \nonumber \\
&\langle 0_{in}|b^{\dagger}_{R,\omega}b_{R,\omega}|0_{in} \rangle \equiv \int^{\infty}_{0}d\omega' |\beta^{RL}_{\omega\omega'}|^{2}, \quad \langle 0_{in}|b^{\dagger}_{L,-\omega}b_{L,-\omega}|0_{in} \rangle \equiv \int^{\infty}_{0}d\omega' |\beta^{LR}_{\omega\omega'}|^{2}\,.
\end{align}

Since we are interested in understanding the nature of particle creation due to (\ref{bc3}), we only need to evaluate $\beta_{\omega\omega'}^{RL} = \beta_{\omega\omega'}^{LR}$. As Klein Gordon inner product is independent of the Cauchy slice we choose, we evaluate the coefficients on $\mathcal{I}^{-}_{R}$. Using (\ref{in}), (\ref{out}) and (\ref{KG}) we have
\begin{align}\label{beta}
\beta_{\omega\omega'}^{RL} &= -(\phi^{out}_{R,\omega},\phi^{in*}_{L,-\omega'})_{K.G,\mathcal{I}^{-}_{R}} \nonumber \\
& = \frac{i}{4\pi\sqrt{\omega\omega'}}\int^{0}_{-\infty}dv\left[-i\omega' e^{-i\omega' v}(e^{-i\omega p^{-1}(v)}+e^{-i\omega u})+i\omega(p^{-1}(v))' e^{-i\omega p^{-1}(v)}(e^{-i\omega' v}+e^{-i\omega'p(u)})\right]_{u=-\infty}
\end{align}
Since the late time behavior of the interface profile gives rise to thermal stress-energy tensor, we approximate the relevant form of the profile as
\begin{align}\label{appr prof}
p(u) \approx -\beta e^{-\frac{u}{\beta}}\,, \quad p^{-1}(v) \approx -\beta \log\left(-\frac{v}{\beta}\right).
\end{align}
In (\ref{beta}), the integrals containing $u$ dependent part vanishes simply due to the fact that $\omega,\omega'>0$. To do the integrals, we first regulate it by using $v \rightarrow v+i\epsilon$ with $\epsilon \rightarrow 0^{+}$ \footnote{A related discussion of the integrals in the context of Unruh radiation is discussed in \cite{Crispino:2007eb}.}. Then, after choosing the integration contour to the negative imaginary axis by substituting $v=-\frac{ix}{\omega'}$ and using (\ref{appr prof}), we have
\begin{align}
\beta_{\omega\omega'}^{RL} &= \frac{e^{-\frac{\beta\pi\omega}{2}}}{4\pi}\left(\frac{1}{\beta}\right)^{i\beta\omega}(\omega')^{-i\beta\omega}\left[\frac{1}{i\omega'}\Gamma(1+i\beta\omega)\sqrt{\frac{\omega'}{\omega}}+\beta\Gamma(i\beta\omega)\sqrt{\frac{\omega}{\omega'}}\right] \nonumber \\
& =  \frac{e^{-\frac{\beta\pi\omega}{2}}}{2\pi}\left(\frac{1}{\beta}\right)^{i\beta\omega-1}(\omega')^{-i\beta\omega}\sqrt{\frac{\omega}{\omega'}}\Gamma(i\beta\omega)
\end{align}
Hence, we can easily see
\begin{align}\label{numb1}
|\beta_{\omega\omega'}^{RL}|^{2} = |\beta_{\omega\omega'}^{LR}|^{2} = \frac{\beta}{4\pi^{2}\omega'}\frac{1}{e^{2\pi\beta\omega}-1}
\end{align}
Remarkably, this gives the exact same number operator expectation value as that of a single moving mirror and of a single-sided 2$D$ black hole formed under null shell collapse\cite{Birrell:1982ix,Hawking:1975vcx,PhysRevD.46.2486}. In a similar fashion, we can obtain the other Bogoliubov coefficient:
\begin{align}
\alpha^{RL}_{\omega\omega'} = \frac{e^{\frac{\beta\pi\omega}{2}}}{2\pi}\left(\frac{1}{\beta}\right)^{-i\beta\omega-1}(\omega')^{i\beta\omega}\sqrt{\frac{\omega}{\omega'}}\Gamma(-i\beta\omega)
\end{align}
 Also, one can see
\begin{align}
|\alpha^{LR}_{\omega\omega'}|^{2} = |\alpha^{RL}_{\omega\omega'}|^{2}
\end{align}
Thus, one could check (\ref{bogo const 1},\ref{bogo const 2}) are satisfied trivially\footnote{Upto a constant normalization factor which is unimportant and can be absorbed in the definition of the coefficients.}. Hence, we end up with
\begin{align}
\langle 0_{in}|b^{\dagger}_{R,\omega}b_{R,\omega}|0_{in} \rangle = \langle 0_{in}|b^{\dagger}_{L,\omega}b_{L,\omega}|0_{in} \rangle = \int^{\infty}_{0}d\omega' \frac{\beta}{4\pi^{2}\omega'}\frac{1}{e^{2\pi\beta\omega}-1}
\end{align}
Hence, the average number operators on both sides at future infinities are the same in $\beta \rightarrow 0$. In other words, both R and L regions are thermal objects having a thermal stress tensor with inverse temperature $\beta$. This concludes a first step toward the construction of a wormhole.\\\\
\textit{\underline{Maximal entanglement and connectedness}:}
We have already constructed the state $|0_{in}\rangle$ defined in $\mathcal{I}^{+}_{L}$ and $\mathcal{I}^{-}_{R}$ from the quantization of field. Similarly we could construct the combined out state as $|0_{out}\rangle$ defined in $\mathcal{I}^{-}_{L}$ and $\mathcal{I}^{+}_{R}$. The particle production in future infinities suggests that the two states will not be the same. Since we compute all the Bogoliubov coefficients, it is a well-known straightforward exercise to write down $|0_{in}\rangle$ in terms of $|0_{out}\rangle$ \cite{Crispino:2007eb}. One can show that in $\beta \rightarrow 0$ limit,
\begin{align}
|0_{in}\rangle = \frac{1}{\sqrt{Z(\beta)}}\sum_{n}e^{-\beta E_{n}}|n_{L}\rangle |n_{R}\rangle
\end{align}
where $|n_{R}\rangle \equiv \frac{1}{\sqrt{n!}}(b^{\dagger}_{\omega,R})^{n_{\omega}}|0_{out}\rangle$ and $|n_{L}\rangle \equiv \frac{1}{\sqrt{n!}}(b^{\dagger}_{-\omega,L})^{n_{\omega}}|0_{out}\rangle$. Thus, the future observer will see a thermal vacuum. This is exactly the TFD state we want to construct. The construction suggests the state is analogous to the thermal Hartle-Hawking state. Hence, it will naturally satisfy all the desired properties of TFD. Though we have studied the free scalar field in the TFD construction, it should be generalized to arbitrary ICFT. We might be able to generalize the construction for two copies of CFTs, CFT$_{L}$ and CFT$_{R}$, separated by interface satisfying boundary condition (\ref{bc3}) in the static frame. The ICFT vacuum state $|I_{h}\rangle$ should be similar to the TFD state. The state satisfying (\ref{bc3}) will be of the form\footnote{For more details see appendix (\ref{app:B}).}
\begin{align}
(T_{L}(\tilde{u})-\bar{T}_{R}(\tilde{v}))|I_{h}\rangle = 0\,, \quad |I_{h}\rangle = \sum_{\vec{k}}|h,\vec{k}\rangle_{L} |\overline{h,\vec{k}}\rangle_{R}\,,
\end{align}
where $|h,\vec{k}\rangle$ is orthonormal basis of the Verma module. However, these states are non-normalizable (infinite norm). One standard way to define a suitable normalizable state is to regularize the state in Euclidean path integral by $e^{-\epsilon(H_{L}+H_{R})}$ with cut-off $\epsilon$. Then the state will be of the form
\begin{align}
|I_{h}\rangle = \frac{1}{\sqrt{Z(\epsilon)}}\sum_{\vec{k}}e^{-\epsilon E(\vec{k})}|h,\vec{k}\rangle_{L} |\overline{h,\vec{k}}\rangle_{R}
\end{align}
where $Z= \sum_{\vec{k}}e^{-\epsilon E(\vec{k})}$ and $E(\vec{k})$ is the total energy at $k$th level. This is very similar to TFD state once we could identify $\epsilon = \beta$. However, this choice of regularization is not unique. To uniquely fix this, we need to construct the state with moving interface conditions like (\ref{bc4}), or we must find the basis of states in the original moving frame. This is an interesting future work to settle this. However, for the purpose of showing the maximal entanglement, the form of $|I_{h}\rangle$ without unique regularization is enough and suggestive.

One could also study the Left-Left(LL) or Right-Right(RR) correlator. Now for the boundary condition (\ref{bc3}), we first use conformal transformation $\tilde{u}=p(u)$ and $\tilde{v}=v$ to write them as correlator in the presence of static interface, and we get
\begin{align}
\langle \mathcal{O}_{R}(u_{1},v_{1}) \mathcal{O}_{R}(u_{2},v_{2}) \rangle_{I,\beta} &=  \left(\frac{\partial \tilde{u}_{1}}{\partial u_{1}}\right)^{h}\left(\frac{\partial \tilde{v}_{1}}{\partial v_{1}}\right)^{h}\left(\frac{\partial \tilde{u}_{2}}{\partial u_{2}}\right)^{h}\left(\frac{\partial \tilde{v}_{2}}{\partial v_{2}}\right)^{h}\langle\mathcal{O}_{L}(\tilde{u}_{1},\tilde{v}_{1}) \mathcal{O}_{R}(\tilde{u}_{2},\tilde{v}_{2}) \rangle_{I} \nonumber \\
& \xrightarrow[\beta \rightarrow 0]{} \beta^{-2h} \sinh(\frac{u_{2}-u_{1}}{2\beta})^{-2h}\frac{1}{(v_{1}-v_{2})^{2h}}
\end{align}
This is similar to the thermal two-point function or LL correlator in TFD case if we take $v=0$ of the probe operators. We can also compute the Left-Right(LR) correlator from the Right-Right(RR) correlator using analytic continuation $t \rightarrow t+i\pi\beta$ as we mentioned earlier. This is analogous to the analytic continuation in time for computing LR correlator in TFD state.
To compute connectedness, we need to compute:
\begin{align}
\langle \mathcal{O}_{L}(u_{1},v_{1}) \mathcal{O}_{R}(u_{2},v_{2}) \rangle_{I,\beta} - \langle \mathcal{O}_{L}(u_{1},v_{1}) \rangle_{I,\beta} \langle \mathcal{O}_{R}(u_{2},v_{2}) \rangle_{I,\beta} 
\end{align}
On the other hand, with the same boundary condition, the purely left(right) )-sided correlators will be unaffected; therefore, in particular, the one-point function will vanish. i.e.,
\begin{align}
\langle \mathcal{O}_{L}(u_{1},v_{1}) \rangle_{I,\beta} &= \left(\frac{\partial \tilde{u}_{1}}{\partial u_{1}}\right)^{h}\left(\frac{\partial \tilde{v}_{1}}{\partial v_{1}}\right)^{h}  \langle\mathcal{O}_{L}(\tilde{u}_{1},\tilde{v}_{1}) \rangle_{I} = 0
\end{align}
Thus, the connected correlator gives a maximum value. This is again consistent with the connected TFD correlator. Hence, the appearance of TFD state, along with the properties of correlators in our setting, provides a one-to-one map between moving interface with two exteriors of eternal BH.\\\\
\textit{\underline{Transmission and Reflection coefficient}:}
In the usual eternal BH settings, the two exteriors are disconnected, and there will be no interaction between them. Even though we get two-sided Hawking radiation, in our setup of oppositely time-directed CFTs separated by a moving interface with boundary condition (\ref{bc3}), the two CFTs seemed to have interacted via the interface. However, here we see this will not be the case when we take $\beta \rightarrow 0$ limit. Here, the reflection coefficients $\mathcal{R}_{L,R}$ are vanishing in this case by definition. To get the transmission coefficients, we again do a scattering experiment as discussed in the last section as in (\ref{modified transport}). In L, the total incoming energy is $\int^{\infty}_{-\infty}dv\bar{T}_{L}(v)$, defined in $\mathcal{I}^{+}_{L}$. To measure transmitted energy $\int^{\infty}_{-\infty}du T_{R}(u)$ and $\int^{\infty}_{-\infty}dv \bar{T}_{R}(v)$, we prepare an anti-chiral state $|\mathcal{\bar{O}}_{L,D}(v)\rangle$. Then the (modified) transmission coefficients in the original moving frame ($u,v$) for boundary condition (\ref{bc3}) corresponding to static ($\tilde{u},\tilde{v}$) frame are defined as (in a similar fashion to (\ref{modified transport})):
\begin{align}
\mathcal{\widetilde{T}}_{L} = \lim_{D\rightarrow \infty} \frac{\langle \mathcal{\bar{O}}_{L,D}| \int^{\infty}_{-\infty}du T_{R}(u)|\mathcal{\bar{O}}_{L,D}\rangle_{I}}{\langle \mathcal{\bar{O}}_{L,D}| \int^{\infty}_{-\infty}dv \bar{T}_{L}(v)|\mathcal{\bar{O}}_{L,D}\rangle} ,\quad\;
\mathcal{\widetilde{T}}_{R} = \lim_{D\rightarrow \infty} \frac{\langle \mathcal{\bar{O}}_{R,D}| \int^{\infty}_{-\infty}du T_{L}(u)|\mathcal{\bar{O}}_{R,D}\rangle_{I}}{\langle \mathcal{\bar{O}}_{R,D}| \int^{\infty}_{-\infty}dv \bar{T}_{R}(v)|\mathcal{\bar{O}}_{R,D}\rangle}
\end{align}
A similar analysis as in the previous section gives: 
\begin{align}
\mathcal{\widetilde{T}}_{L} = \frac{2}{1+e^{\frac{\ell}{\beta}}}, \; \mathcal{\widetilde{T}}_{R} = \frac{2}{1+e^{\frac{\ell}{\beta}}}, \; \mathcal{T}_{L}=\mathcal{T}_{R}=0
\end{align}
In $\beta \rightarrow 0$ limit, we get $\mathcal{\widetilde{T}}_{R,L} \rightarrow 0$. This suggests the interface subjected to (\ref{bc3}) becomes non-interacting when $\beta \rightarrow 0$. This is consistent with the two exteriors of eternal BH spacetime.

\section{Discussions}\label{sec:discussion}
\textit{\underline{Summary}:}
In this paper, we have generalized the idea of `radiative BCFT' introduced in \cite{Akal:2021foz} to a broader class of boundary conditions which we rephrase as `radiative ICFT'. The crucial feature of radiative boundary conditions is that the right and left-moving energy-momentum tensors of those boundaries are not on equal footing. This, in turn, provides a non-vanishing energy flux at the boundary. In our construction, we consider two different gluing conditions for ICFT constructed out of two CFTs interacting at the radiative interface. For the ICFT of $CFT_{L}$ and $CFT_{R}$, we have found that $\mathcal{R}+\mathcal{T} \neq 1$ in each side. In particular, we have $(\mathcal{R},\mathcal{T})<1$ for any finite $\beta$. However, at $\beta \rightarrow \infty$\footnote{Eventually, that mimics $u \rightarrow -\infty$ limit when the dynamical interface approaches a static one.}, we will always have $\mathcal{R}+\mathcal{T} =1$. On the other hand, the departure of $\mathcal{R}+\mathcal{T}$ from one is maximum when we take $\beta \rightarrow 0$ limit when the interface approaches a null line. In this limit,  the interface behaves like a `semipermeable membrane'.

These observations suggest that the dynamical profile of the interface is solely responsible for the absorption or radiation of energy for whatever boundary conditions we impose. On the other hand, if we consider the ICFT boundary conditions for $CFT_{L}$ and $\overline{CFT}_{R}$ where the net energy flux of one side is exactly the same but negative with respect to the other side, the interface behaves like a horizon (for totally transmitive solution) in $\beta \rightarrow 0$ limit. In this case, if we perform a modified scattering experiment,
we find that the $CFT_{R}$ is completely frozen with no transmission or reflection while the $CFT_{L}$ purely transmits energy from L to R. This feature motivated us to model an eternal black hole (or wormhole) like setting. In this context, we have argued when we consider ICFT subjected to the same boundary condition (\ref{bc3}) with an upward time direction in R and downward time direction in L, the average particle production in each side is similar to that in the exteriors of eternal black hole.  This, in turn, makes the vacuum of the ICFT as the TFD state, which has the defining properties of a wormhole state following ER=EPR. By computing transmission coefficients, we have shown that both copies are actually non-transmittive in the same limit. Based on our observations, we will now make some comments and speculative remarks that, we feel, might provide some potential future directions worth exploring.\\\\
\textit{\underline{Evolution of interior}:}
Using the gluing condition (\ref{bc3}), we have shown that we may interpret two CFTs as models of exterior-interior or exterior-exterior of an eternal black hole by showing thermal particle production in the exterior as well as by studying scattering experiments defined naturally in these settings. We may think of this as a CFT toy version of ER=EPR. However, we could argue that the growth of a wormhole or Einstein-Rosen bridge has an implicit structural similarity to the transmitivity of moving interface with different boundary conditions.


 We first summarize the interface boundary conditions (at $\tilde{u}=\tilde{v}$) of $CFT_{L}\otimes \overline{CFT}_{I}$ and the corresponding reflection and transmission coefficients we have studied so far in the last section.
 \begin{itemize}
\item \textit{Gluing condition I}: \begin{align}\label{a1}
&T_{L} = \bar{T}_{I}, \; \bar{T}_{L}=T_{I}, \\
&\mathcal{R}_{L}=\mathcal{R}_{I} = 0; \; \mathcal{\widetilde{T}}_{L}=1, \; \mathcal{\widetilde{T}}_I \xrightarrow[\beta\rightarrow 0]{} 0\,.
\end{align}
The one-sided transmitivity of the interface reflects the classical smooth horizon-like property of eternal black hole geometry. This boundary condition is reminiscent of the usual ingoing boundary condition for a smooth horizon.

\item \textit{Gluing condition II}: \begin{align}\label{a2}
&T_{L}+T_{I} = \bar{T}_{I}+\bar{T}_{L}, \\
&\mathcal{R}_{L}= 1-\frac{c_{L\bar{I}}}{c_{L}},\; \mathcal{R}_{I} \xrightarrow[\beta\rightarrow 0]{} 0; \; \mathcal{\widetilde{T}}_{L}=\frac{c_{L\bar{I}}}{c_{L}}, \; \mathcal{\widetilde{T}}_{I} \xrightarrow[\beta\rightarrow 0]{} 0
\end{align}
(\ref{a2}) is the most general conformal boundary condition for $CFT_{L}\otimes \overline{CFT}_{I}$ and (\ref{a1}) is one special solution of that. Since $\mathcal{\widetilde{T}}_{L}<1$, the smoothness of the interface, aka the horizon (as defined through the previous case) is modified. The non-zero $R_{L}$ mimics that there is a possibility of a structure at the horizon that has no classical general relativity analog. The I region remains unaffected, though. 
\item \textit{Gluing condition III}:
\begin{align}\label{a3}
&T_{L} = \bar{T}_{L}, \; \bar{T}_{I}=T_{I}, \\
&\mathcal{R}_{L}=1, \; \mathcal{R}_{I} \xrightarrow[\beta\rightarrow 0]{} 0; \; \mathcal{\widetilde{T}}_{L}=0, \; \mathcal{\widetilde{T}}_{I} \xrightarrow[\beta\rightarrow 0]{} 0
\end{align}
For this condition, we must have $c_{\bar{L}I}=0$. This condition reflects the transmitivity from the left side is completely lost in the relevant $\beta \rightarrow 0$ limit. The interface behaves like an opaque brick wall. L and I regions become disconnected and unentangled. Again, this has no analogous semiclassical gravity description.
\end{itemize}
In the context of an eternal black hole, we know spatial slices inside the interior grow in time. This time dependence of the interior is a remarkable feature even when the black hole is in thermal equilibrium \cite{Susskind:2014rva,Susskind:2014moa,Susskind:2018pmk}. The growth of those slices is linear in time, as expected classically. However, after a long time, the volume of the ER bridge approaches its maximum value, and the growth saturates there. This is expected to be around a very large time scale $t \sim e^{S}$ \cite{Susskind:2014rva,Susskind:2014moa}. The bound of this growth is purely a quantum effect, which provides a time scale where the classical general relativity breaks down. The classical linear growth, as well as quantum mechanical saturation, have a striking similarity to the computational complexity\cite{Susskind:2014rva,Susskind:2014moa,Susskind:2018pmk}. Another way to understand this growth is by studying connected LR correlators in the eternal black hole background. Initially, at $t=0$ the correlator has its maximum value. As time increases, it decreases exponentially in time. It is expected when the wormhole length attains its maximum value, the correlator vanish. \\
We will now return to our observation to connect wormhole-like physics in the presence of a moving interface. From the observation of (\ref{a1})-(\ref{a3}), we can speculate a similar story in our model for $CFT_I$. The total (time-averaged) volume of I region in $\beta \rightarrow 0$ limit will be proportional to net energy accumulation from L to I subjected to different boundary conditions. In other words, we can say that the quantity $(\mathcal{\widetilde{T}}_{L}-\mathcal{\widetilde{T}}_{I})$ characterizes the net energy flow from L to I. 

From (\ref{a1}), this implies a linear constant growth of I region. (\ref{a2}) suggests a slight decrease of growth rate since $\mathcal{\widetilde{T}}_{L}-\mathcal{\widetilde{T}}_{I}=\frac{c_{L\bar{R}}}{c_{L}}<1$. In (\ref{a3}), the quantity becomes zero, and hence the I region becomes stagnant. Note that the sequence (\ref{a1})$\rightarrow$(\ref{a2})$\rightarrow$(\ref{a3}) is self-consistent from the point of view of changing boundary conditions from one to another. One can not tune (\ref{a1}) to (\ref{a3}) without passing through (\ref{a2}). For example, consider the following intermediate semi-transparent boundary condition \footnote{We want to thank Bobby Ezhuthachan for pointing out this interesting observation.}
\begin{align}\label{alpha}
T_{L}=\alpha \bar{T}_{L} + (1-\beta) \bar{T}_{I}
\end{align}
A similar condition can be applied for $T_{I}$ with some coefficients. From (\ref{alpha}), we could see when $\alpha=\beta=0$, we can tune to (\ref{a1}). After increasing $\alpha,\beta$, we can have a similar structure of (\ref{a2}). When we reach $\alpha=\beta=1$ we have (\ref{a3}). Thus, changing $\alpha,\beta$ from 0 to 1 gives a sequence of boundary conditions in the same way from (\ref{a1}) to (\ref{a3}). Hence, we see that the sequence of changing boundary conditions has a one-to-one correspondence with the wormhole growth at different epochs of time.\\
It would be an interesting future problem to make all these claims more concrete by studying complexity in the presence of a moving interface\footnote{In moving mirror-black hole connection, subregion complexity has been studied in \cite{Sato:2021ftf}.}. The ground states of the ICFT under different boundary conditions will be different. Hence, to construct the state corresponding to (\ref{a3}) from (\ref{a1}), we need to use multiple unitary gates(here operators) that tune $\alpha,\beta$ of (\ref{alpha}). Similarly, we could also study connected LR correlator for the sequence of boundary conditions from (\ref{a1}) to (\ref{a3}). While for (\ref{a1}) it is maximally connected, in (\ref{a3}) it is completely disconnected. We hope to return more concretely in this direction in the near future. \\\\
\textit{\underline{Holography}:}
One interesting future study is to visualize our construction in holography. Recently, authors of \cite{Bachas:2020yxv,Bachas:2022etu} have computed energy reflection and transmission coefficients using thin-brane holographic duals, which consists of two AdS$_3$ slices joined by a brane having tension $\sigma$. The bulk junction condition is enough to capture the ICFT transmission and reflection coefficient subjected to a
CFT scattering experiment. Transmission and reflection coefficients depend on the radius of the AdS$_3$ slices, which is related to the central charges $c_L$ and $c_R$ of two CFTs through the Brown-Henneaux formula \cite{Brown:1986nw}, it also depends on the tension of the brane $\sigma$. On the other hand, in the CFT computation \cite{Meineri:2019ycm}, transmission and reflection coefficients depend on central charges $c_L$ and $c_R$, as well as on $c_{LR}$ defined through the expectation value of left and right stress tensor. The holographic computations, therefore, give a gravitational derivation of this CFT data.  It would be an interesting problem to generalize their method \cite{Bachas:2020yxv,Bachas:2022etu} in our case, where we hope to get a genuine match as all the CFT data have already been fixed in terms of gravitational parameters. However, a more direct bulk scattering experiment(by sending shockwaves from the boundary) dual to CFT one is itself an interesting area to look at. Apart from that, to materialize the BH connection in the bulk, it would be very interesting to construct the dual corresponding to (\ref{bc3}). This might tell exactly how to create a real interior by merging two different spacetimes using dynamical branes. As of now, we are working on an analogue CFT toy version. To address most of the interior geometry as well as bulk scattering experiments, holography could be the best way to look at it. We hope to return to address some of these in the near future. It would also be important to understand evaporating black holes in our framework. For that, we may need to study the kink interface profile \cite{Akal:2021foz} and its gravitational dual.\\\\
\textit{\underline{Comments about moving mirror-black hole connection}:}
In the original moving mirror-black hole connection picture, the horizon remained obscure due to the lack of an independent definition. In the one-sided picture, $\tilde{u}=0$ or $u = \infty$ was considered to be the horizon. In our case, we see at $\beta \rightarrow 0$ limit, this eventually coincides. However, in moving mirror, there is no region beyond the horizon, aka interior. Nevertheless, the analogy was still striking due to the evidence of Hawking radiation at the null future boundary as well as the thermal stress tensor profile. In our framework (for instance, see (\ref{a3})), we find pieces of evidence that when the interface becomes a one-sided mirror, the smoothness of the horizon(aka the interface) goes away. This becomes an opaque boundary. A recent study in \cite{Burman:2023kko} also finds pieces of evidence that in a Planckian stretched horizon (brick-wall) framework; one can still get thermal Hawking radiation(up to exponentially small correction in entropy) in spite of the lack of smoothness. This was eventually a model for typical states\footnote{In \cite{Banerjee:2024dpl}, some aspects of `thermality' in Green's function of stretched horizon background have been studied alternatively.}. From the geometric point of view, the transition from transparency to opaqueness of the boundary condition (for instance, (\ref{a1}) to (\ref{a3})) suggests a dramatic change in the smoothness of the horizon. Our results are very indicative of the plausibility of firewall in typical states\footnote{Here what we mean by typical states are those geometries formed by collapse.} \cite{Marolf:2013dba}. We have some evidences that the bulk dual of the moving mirror is very similar to brick wall geometry where the ETW brane itself behaves like a horizon. Hence, from our moving interface results, we are tempted to view the moving mirror as a typical state or state created by natural processes like collapse having a firewall rather than a mimicker of an eternal black hole. We hope to return with some interesting results soon to make this discussion concrete.
%

\section{Acknowledgements}
We would like to thank Vaibhav Burman, Justin David, Bobby Ezhuthachan, Dongsheng Ge, Chethan Krishnan, Somnath Porey, Koushik Ray, Baishali Roy and Tadashi Takayanagi for useful related discussions. We would like to thank Baishali and Somnath for collaborating with us at the initial stage of the project. We are especially grateful to Bobby for introducing us to ICFT as well as for suggesting us one of the key problems of this project. We thank him for his constant support and encouragement throughout this work and for clarifying some key points of this work. We also thank him for reading our draft carefully and suggesting some conceptual changes therein. We would also like to thank the Physics Department of RKMVERI as well as the organizers of the Workshop on ‘Ergodicity and its breaking: A view from Many Body, QFT and Holography’ at RKMVERI, where this project was discussed and initiated. AD and SD would like to thank the hospitality of the Physics department, RKMVERI while visiting there in the course of this project. We would also like to thank Chethan Krishnan for the course on `Quantum Black Holes' \cite{youtube} at CHEP, IISC, which proved to be helpful and illuminating in getting a key motivation on some part of the project. AD would like to thank Tadashi Takayanagi for useful in-person discussion during The Kavli Asian Winter School, 2023, in Kyoto and he would also like to thank the organizers of the school for giving him an opportunity to present some part of this work. PB would like to acknowledge the support provided by the grant CRG/2021/004539. The research work of SD is supported by a DST Inspire Faculty Fellowship.

\appendix
\section{Details of computing translation and reflection coefficients}\label{A1}
In this appendix, we will give details of computations of the transmission coefficient (eq.\eqref{main1}) and reflection coefficient  (eq.\eqref{main2}).
\subsection{For $CFT_{L}$} 
\textit{\underline{The Transmission Coefficient}:}
Firstly we will do the $\Tilde{u}$ integration of $\braket{O_{L},D|\mathcal{E}_{R}|O_{_{L}},D}_{I}$ , and leave the other integrations for later. $\braket{O_{L},D|\mathcal{E}_{R}|O_{_{L}},D}_{I}$ without the $\Tilde{u}_{1}$ and $\Tilde{u}_{2}$ integration is

\begin{equation}\label{eq:int}
\begin{split} 
&-\frac{h}{2 \pi} \frac{c_{LR}}{c_{L}}\int_{-\infty}^{0} d\Tilde{u} \frac{(1-e^{\Tilde{u}/\beta})^{2}}{1-e^{\Tilde{u}/\beta}} \Bigg{[}\frac{1}{(\Tilde{u}-\Tilde{u}_{1})^{2}(\Tilde{u}_{1}-\Tilde{u}_{2})^{2h}}+\frac{1}{(\Tilde{u}-\Tilde{u}_{2})^{2}(\Tilde{u}_{1}-\Tilde{u}_{2})^{2h}}\\
&-\frac{2}{(\Tilde{u}-\Tilde{u}_{1})(\Tilde{u}_{1}-\Tilde{u}_{2})^{2h+1}}-\frac{2}{(\Tilde{u}-\Tilde{u}_{2})(\Tilde{u}_{1}-\Tilde{u}_{2})^{2h+1}}\Bigg{]}\\
=& -\frac{h}{2 \pi} \frac{c_{LR}}{c_{L}}
\int_{-\infty}^{0} d\Tilde{u} \Bigg{[}\frac{(1-e^{\Tilde{u}/\beta}) }{(\Tilde{u}-\Tilde{u}_{1})^{2}(\Tilde{u}_{1}-\Tilde{u}_{2})^{2h}}-\frac{2h(1-e^{\Tilde{u}/\beta}) }{(\Tilde{u}-\Tilde{u}_{1})(\Tilde{u}_{1}-\Tilde{u}_{2})^{2h+1}}\Bigg{]} + (\Tilde{u}_{1} \Longleftrightarrow \Tilde{u}_{2} )\\
=&-\frac{h}{2 \pi} \frac{c_{LR}}{c_{L}}\Bigg(\frac{A}{(\Tilde{u}_{1}-\Tilde{u}_{2})^{2h}}-\frac{2B+2C}{(\Tilde{u}_{1}-\Tilde{u}_{2})^{2h+1}} \Bigg)  + (\Tilde{u}_{1} \Longleftrightarrow \Tilde{u}_{2} )
\end{split}
\end{equation}
Where,
\begin{equation}
\begin{split}
A &= \int_{-\infty }^{\tilde{u}_{1}-\epsilon } \frac{1-e^{\tilde{u}/\beta }}{(\tilde{u}-\tilde{u}_{1})^2} \, d\tilde{u}+\int_{\tilde{u}_{1}+\epsilon }^0 \frac{1-e^{\tilde{u}/\beta }}{(\tilde{u}-\tilde{u}_{1})^2} \, d\tilde{u}\\ & =\frac{\beta -\epsilon  e^{\tilde{u}_{1}/\beta } \text{Ei}\left(-\frac{\epsilon }{\beta }\right)+\beta  \left(-e^{\frac{\tilde{u}_{1}+\epsilon }{\beta }}\right)+\beta  \left(1-e^{\frac{\tilde{u}_{1}-\epsilon }{\beta }}\right)+\epsilon  e^{\tilde{u}_{1}/\beta } \left(\Gamma \left(0,\frac{\tilde{u}_{1}}{\beta }\right)-\Gamma \left(0,-\frac{\epsilon }{\beta }\right)\right)}{\beta  \epsilon } \\
B &=\int_{-\infty }^{\tilde{u}_{1}-\epsilon } \frac{e^{\tilde{u}/\beta }}{\tilde{u}-\tilde{u}_{1}} \, d\tilde{u}+\int_{\tilde{u}_{1}+\epsilon }^0 \frac{e^{\tilde{u}/\beta }}{\tilde{u}-\tilde{u}_{1}} \, d\tilde{u} \\& =e^{\tilde{u}_{1}/\beta } \text{Ei}\left(-\frac{\epsilon }{\beta }\right)+e^{\tilde{u}_{1}/\beta } \left(\Gamma \left(0,-\frac{\epsilon }{\beta }\right)-\Gamma \left(0,\frac{\tilde{u}_{1}}{\beta }\right)\right) \\ 
C&=\int_{-\frac{1}{\epsilon }}^{\tilde{u}_{1}-\epsilon } \frac{1}{\tilde{u}-\tilde{u}_{1}} \, d\tilde{u}+\int_{\tilde{u}_{1}+\epsilon }^0 \frac{1}{\tilde{u}-\tilde{u}_{1}} \, d\tilde{u} \\&=\log \left(-\frac{\tilde{u}_{1} \epsilon }{\tilde{u}_{1} \epsilon +1}\right),\quad\text{if }\Big(\tilde{u}_{1}+\frac{1}{\epsilon }\Big)>\epsilon 
\end{split}
\end{equation}
In the limit $\epsilon \rightarrow 0$, we have the followings
\begin{equation}\label{eq:limit}
\begin{split}
\lim_{\epsilon \to 0} \epsilon A = 2-2 e^{\tilde{u}_{1}/\beta }\,, \hspace{1cm} \lim_{\epsilon \to 0} &\epsilon B= 0\,,\hspace{1cm} \lim_{\epsilon \to 0} \epsilon C=0\,.
\end{split}
\end{equation}
Using \eqref{eq:limit}, we get the final answer for the integration \eqref{eq:int} to be
\begin{equation}
-\frac{h}{2 \pi} \frac{c_{LR}}{c_{L}} \frac{ 1}{(\Tilde{u}_{1}-\Tilde{u}_{2})^{2h}} \Bigg[ 4-2\,e^{\Tilde{u}_{1}/\beta}-2\,e^{\Tilde{u}_{2}/\beta}\Bigg] \frac{1}{\epsilon} \,.
\end{equation}
\textit{\underline{The denominator}:}
Let's look at the denominator after a generic transformation.
\begin{equation}
\begin{split}
&\braket{O_{L},D|\mathcal{E}_{L}|O_{_{L}},D} \\= & -\frac{h}{2 \pi} \int_{-\infty}^{\infty}du\, du_{1}du_{2} 
f(u_{1})f(u_{2}) \Bigg{[}\frac{1}{(u-u_{1})^{2}(u_{1}-u_{2})^{2h}}-\frac{2}{(u-u_{1})(u_{1}-u_{2})^{2h+1}}+(\Tilde{u}_{1} \Longleftrightarrow \Tilde{u}_{2} )\Bigg{]}
\end{split}
\end{equation}
Like the numerator, we will evaluate the $u$-integration first. 
\begin{equation}
\begin{split}
 & -\frac{h}{2 \pi} \int_{-\infty}^{\infty}du  \Bigg{[}\frac{1}{(u-u_{1})^{2}} -\frac{2}{(u-u_{1})}+(\Tilde{u}_{1} \Longleftrightarrow \Tilde{u}_{2} )\Bigg{]}\\ &  
\end{split}
\end{equation}
The first term of the  $u$-integration 
\begin{equation}
\begin{split}
&\frac{h}{(u_{1}-u_{2})^{2h}} \int^{\infty}_{-\infty} \frac{du}{(u-u_{1})^{2}} \hspace{.5cm} \text{diverges at $u=u_{1}$} \\ &=\lim_{\epsilon \to 0} \frac{h}{(u_{1}-u_{2})^{2h}} \Bigg{[} \int_{-\infty}^{u_{1}-\epsilon} \frac{du}{(u-u_{1})^{2}}+ \int^{\infty}_{u_{1}+\epsilon} \frac{du}{(u-u_{1})^{2}} \Bigg{]}\\&==\lim_{\epsilon \to 0} \frac{h}{(u_{1}-u_{2})^{2h}} \Bigg{[} -\frac{1}{u_{1}-\epsilon-u_{1}}+\frac{1}{u_{1}+\epsilon-u_{1}} \Bigg{]} \\& =\lim_{\epsilon \to 0} \frac{h}{(u_{1}-u_{2})^{2h}} \frac{2}{\epsilon}
\end{split}
\end{equation}
The second term of the $u$-integration

\begin{equation}
\begin{split}
&-\frac{2h}{(u_{1}-u_{2})^{2h+1}} \int ^{\infty}_{-\infty} \frac{du}{u-u_{1}} \\ &= \lim_{\substack{\Lambda \to \infty \\ \epsilon \to 0} }-\frac{2h}{(u_{1}-u_{2})^{2h+1}}  \Bigg{[} \int_{-\Lambda}^{u_{1}-\epsilon}   \frac{du}{u-u_{1}}  +  \int^{\Lambda}_{u_{1}+\epsilon}   \frac{du}{u-u_{1}} \Bigg]\\  &= \lim_{\substack{\Lambda \to \infty \\ \epsilon \to 0} }-\frac{2h}{(u_{1}-u_{2})^{2h+1}}  \Bigg{[}\log{\frac{u_{1}-\epsilon-u_{1}}{u_{1}+\epsilon-u_{1}}}-\log{\frac{-\Lambda-u_{1}}{\Lambda-u_{1}}} \Bigg] \\ &=0
\end{split}
\end{equation}
So, the denominator after doing the $u$-integration is
\begin{equation}
\begin{split}
\braket{O_{L},D|\mathcal{E}_{L}|O_{_{L}},D}= \lim_{\epsilon\to 0} -\frac{1}{2 \pi} \int du_{1} du_{2} f(u_{1}) f(u_{2}) \Bigg[\frac{2}{\epsilon} \frac{h}{(u_{1}-u_{2})^{2h}}\Bigg]
\end{split}
\end{equation}
So, the final expression after only evaluating the $u$-integration  is 
\begin{equation}
\begin{split}
\mathcal{T}_{L }&=\lim_{D\to\infty}\frac{\int_{-\infty}^{\infty}du\braket{O_{L}^{1}(u_{1},D)|T_{R}(u)|O_{L}^{2}(u_{2},D)}_{I}}{\int_{-\infty}^{\infty}du\braket{O_{L}^{1}(u_{1},D)|T_{L}(u)|O_{L}^{2}(u_{2},D)}} \\ &= \frac{c_{LR}}{c_{L}} \frac{1}{\int du_{1} du_{2} f(u_{1})f(u_{2})\frac{2h}{(u_{1}-u_{2})^{2h}}} \\ & \times \int d\Tilde{u}_{1}d\Tilde{u}_{2} f(p^{-1}(\Tilde{u}_{1})) f(p^{-1}(\Tilde{u}_{2})) \Bigg(\frac{\partial \Tilde{u}_{1}}{\partial u_{1}}\Bigg)^{h_{1}-1} \Bigg(\frac{\partial \Tilde{u}_{2}}{\partial u_{2}}\Bigg)^{h_{2}-1} \frac{2h}{(\Tilde{u}_{1}-\Tilde{u}_{2})^{2h}} \Bigg[ 4-2e^{\Tilde{u}_{1}/\beta}-2e^{\Tilde{u}_{2}/\beta}\Bigg] 
\end{split}
\end{equation}
Notice that the $\epsilon$ terms cancel each other from the numerator and the denominator. 
Now we are going to do the  $\Tilde{u}_{1}$ and $\Tilde{u}_{2}$ integration. We finally chose some profiles for the states.

\begin{equation}
\begin{split}
f(u)=\delta(u-\ell), \hspace{.5cm} u=\ell \hspace{.1cm}  \longrightarrow \Tilde{\ell}= -\beta \log(1+e^{-\ell/\beta})
\end{split}
\end{equation}
Now we are doing the integral, taking $\ell_{1}$ and $\ell_{2}$ into account. 
\begin{equation}
\begin{split}
f(u_{1})=\delta (u_{1}-\ell_{1}) \hspace{.5cm} \Tilde{\ell}_{1}=- \beta \log(1+e^{-\ell_{1}/\beta}) \\ f(u_{2})=\delta (u_{2}-\ell_{2}) \hspace{.5cm} \Tilde{\ell}_{2}=- \beta \log(1+e^{-\ell_{2}/\beta})
\end{split}
\end{equation}

\begin{equation}
\begin{split}
\mathcal{T}_{L }&=\lim_{D\to\infty}\frac{\int_{-\infty}^{\infty}du\braket{O_{L}^{1}(u_{1},D)|T_{R}(u)|O_{L}^{2}(u_{2},D)}_{I}}{\int_{-\infty}^{\infty}du\braket{O_{L}^{1}(u_{1},D)|T_{L}(u)|O_{L}^{2}(u_{2},D)}} \\ &= \frac{c_{LR}}{c_{L}} \frac{1}{\int_{-\infty}^{\infty} du_{1} \int_{-\infty}^{\infty} du_{2} \delta(u_{1}-\ell_{1})\delta(u_{2}-\ell_{2})\frac{2h}{(u_{1}-u_{2})^{2h}}} \\ & \times \int_{-\infty}^{0}  d\Tilde{u}_{1}  \int_{-\infty}^{0} d\Tilde{u}_{2} \delta(\Tilde{u}_{1}+\beta \log(1+e^{-\ell_{1}/\beta})) \delta(\Tilde{u}_{2}+\beta \log(1+e^{-\ell_{2}/\beta})) \Bigg(1-e^{\Tilde{u}_{1}/\beta}\Bigg)^{h_{1}} \\ & \times \Bigg(1-e^{\Tilde{u}_{1}/\beta}\Bigg)^{h_{2}}  \frac{2h}{(\Tilde{u}_{1}-\Tilde{u}_{2})^{2h}} \Bigg[ 4-2e^{\Tilde{u}_{1}/\beta}-2e^{\Tilde{u}_{2}/\beta}\Bigg] 
\end{split}
\end{equation}
Here, we can easily take the limit $D \rightarrow \infty$ as there is no anti-chiral part. After using $\delta $ functions, and also $h=h_{1}=h_{2}$, we have arrived at 
\begin{equation}
\begin{split}
\mathcal{T}_{L }&=\frac{c_{LR}}{c_{L}} (\ell_{1}-\ell_{2})^{2h} \Bigg(\frac{e^{-\frac{\ell_{1}}{\beta}}}{1+e^{-\ell_{1}/\beta}}\Bigg)^{h} \times\Bigg(\frac{e^{-\frac{\ell_{2}}{\beta}}}{1+e^{-\ell_{2}/\beta}}\Bigg)^{h}  \times \frac{1}{\Big(\beta \log\Big(\frac{1+e^{-\ell_{2}/\beta}}{1+e^{-\ell_{1}/\beta}}\Big)\Big)^{2h}} \\&\times \frac{e^{-\ell_{1}/\beta}+2e^{-\frac{\ell_{1}+\ell_{2}}{\beta}}+e^{-\ell_{2}/\beta}}{(1+e^{-\ell_{1}/\beta})(1+e^{-\ell_{2}/\beta})} \\
\mathcal{T}_{L }&=\frac{c_{LR}}{c_{L}} \Bigg(\frac{\ell_{1}-\ell_{2}}{\beta}\Bigg)^{2h} \frac{1}{\Big( \log\Big(\frac{1+e^{-\ell_{2}/\beta}}{1+e^{-\ell_{1}/\beta}}\Big)\Big)^{2h}} \\&\times \frac{e^{-h(\frac{\ell_{1}+\ell_{2}}{\beta})}(e^{-\ell_{1}/\beta}+2e^{-\frac{\ell_{1}+\ell_{2}}{\beta}}+e^{-\ell_{2}/\beta})}{(1+e^{-\ell_{1}/\beta})^{h+1}(1+e^{-\ell_{2}/\beta})^{h+1}}
\end{split}
\end{equation}
If we take the limit $\ell_{1}\longrightarrow \ell_{2}=\ell$ , we get  the final expression for the transmission coefficient
\begin{equation}
\begin{split}
\mathcal{T}_{L } &= \frac{c_{LR}}{c_{L}}e^{-\frac{2h\ell}{\beta}} (e^{-\frac{\ell}{\beta}})^{-2h}\frac{2}{(1+e^{\frac{\ell}{\beta}})} \\ &= \frac{c_{LR}}{c_{L}}\Bigg( \frac{2}{1+e^{\frac{\ell}{\beta}}} \Bigg)
\end{split}
\end{equation}

\subsection{For $CFT_{R}$}

The same computation can be done for the right $CFT_{R}$. For this case, the states are created by anti-chiral operators belonging to $CFT_{R}$

\begin{equation}
\begin{split}
\mathcal{T}_{R }&=\lim_{D\to\infty}\frac{\int_{-\infty}^{\infty}dv\braket{\bar{O}_{R}^{1}(v_{1},D)|\bar{T}_{L}(v)|\bar{O}_{R}^{2}(v_{2},D)}_{I}}{\int_{-\infty}^{\infty}dv\braket{\bar{O}_{R}^{1}(v_{1},D)|\bar{T}_{R}(v)|\bar{O}_{R}^{2}(v_{2},D)}}
\end{split}
\end{equation}

When defining the states of $CFT_{R}$, the dependency of $D$ will be on $u$, not on $v$, so we can take the $D\rightarrow \infty$ limit. The calculations are straightforward. We will get different results compared to the above calculations as we have transformed only the $u$ coordinates, not the $v$. Following the same steps described above, we get
\begin{equation}
\begin{split}
\mathcal{T}_{R } = \frac{c_{LR}}{c_{R}},  \hspace{1cm} \mathcal{R}_{R } =  \frac{c_{R}-c_{LR}}{c_{R}} \Bigg( \frac{2}{1+e^{\frac{\ell}{\beta}}} \Bigg)\,, 
\end{split}
\end{equation}
\begin{equation}
\begin{split}
\mathcal{T}_{R}+\mathcal{R}_{R}= \frac{2}{1+e^{\frac{\ell}{\beta}}} +\frac{c_{LR}}{c_{R}}\Bigg( \frac{e^{\frac{\ell}{\beta}}-1}{e^{\frac{\ell}{\beta}}+1} \Bigg)\,.
\end{split}
\end{equation}

\section{A detail study of (\ref{bc3})}\label{app:B}

Here, we discuss everything in Euclidean 2$D$ manifold($u \equiv x+i\tau, v \equiv x-i\tau$) to use the powerful complex analysis structure of 2$D$ CFT. All of these Euclidean results can be generalized to Lorentzian after suitable analytic continuation, which we used in the main text.\\\\
\textit{\underline{Ward identity}:}
Consider a tensor product CFT $(T_{L}\otimes \mathbb{1}_{R} +\mathbb{1}_{L}\otimes T_{R})$. We would like to understand Ward identity for mixed correlator $X(u_{1},v_{1},u_{2},v_{2}) \equiv \langle\mathcal{O}^{(h_{1},\bar{h}_{1})}_{L}(u_{1},v_{1})\mathcal{O}_{R}^{(h_{2},\bar{h}_{2})}(u_{2},v_{2})\rangle$ satisfying the boundary condition
\begin{align}
T_{L}(u) + T_{R}(u) = \bar{T}_{L}(v)+\bar{T}_{R}(v), \; \text{at} \; u=v
\end{align}
For infinitesimal conformal transformation $u \rightarrow u+\epsilon(u)$ and $v\rightarrow v+\bar{\epsilon}(v)$, the general conformal Ward identity for the tensor product CFT is
\begin{align}\label{ward1}
&\delta_{\epsilon_{1},\epsilon_{2},\bar{\epsilon}_{1},\bar{\epsilon}_{2}} X  \nonumber \\
&= \frac{1}{2\pi i}\oint_{UHP} du \langle\left(\epsilon_{1}(u)T_{L}(u)+\epsilon_{2}(u)T_{R}(u)\right)X\rangle - \frac{1}{2\pi i}\oint_{UHP} dv \langle\left(\bar{\epsilon}_{1}(v)\bar{T}_{L}(v)+\bar{\epsilon}_{2}(v)\bar{T}_{R}(v)\right)X\rangle 
\end{align}
Here, the subscript $UHP$ refers to the contour of the integral to be the upper half-plane. We will now consider a special class of solution as in (\ref{bc3}). In Euclidean notation, this implies
\begin{align}\label{spcl bc}
T_{L}(u)=\bar{T}_{R}(v), \; \bar{T}_{L}(v) = T_{R}(u); \; \text{at} \; u=v 
\end{align}
One need to further specify the relation between $(\epsilon_{i},\bar{\epsilon}_{i})$ to be consistent with the boundary condition:
\begin{align}\label{epsilon}
\epsilon_{1}(u) = \bar{\epsilon}_{2}(v), \; \bar{\epsilon}_{1}(v) = \epsilon_{2}(u); \; \text{at} \; u=v. 
\end{align}
Using (\ref{epsilon}) in (\ref{ward1}), we get
\begin{align}
&\delta_{\epsilon_{1},\epsilon_{2}} X \nonumber \\
&=  \frac{1}{2\pi i}\oint_{UHP} du \langle\left(\epsilon_{1}(u)T_{L}(u)+\epsilon_{2}(u)T_{R}(u)\right)X\rangle - \frac{1}{2\pi i}\oint_{UHP} dv \langle\left(\epsilon_{2}(v)\bar{T}_{L}(v)+\epsilon_{1}(v)\bar{T}_{R}(v)\right)X\rangle 
\end{align}
Now along the real line $(u=v)$, all the above two terms will be canceled due to (\ref{spcl bc}). Then, using the doubling trick, we can rewrite the UHP integral to a closed contour over full complex plane as the following:
\begin{align}
&\delta_{\epsilon_{1},\epsilon_{2}} X \nonumber \\
&=  \frac{1}{2\pi i}\oint_{c} du \epsilon_{1}(u)\langle \mathbb{T}_{1}(u)X \rangle + \frac{1}{2\pi i}\oint_{c} du \epsilon_{2}(u)\langle \mathbb{T}_{2}(u) X\rangle 
\end{align}
where $\mathbb{T}_{1,2}$ is defined as
\begin{align}\label{T1}
\mathbb{T}_{1}(u) =& T_{L}(u) \; \text{when} \; Im(u)>0, \nonumber \\
& \bar{T}_{R}(v) \; \text{when} \; Im(u)<0.
\end{align}
\begin{align}\label{T2}
\mathbb{T}_{2}(u) =& T_{R}(u) \; \text{when} \; Im(u)>0, \nonumber \\
& \bar{T}_{L}(v) \; \text{when} \; Im(u)<0.
\end{align}
Hence the two completely independent conformal transformation $\epsilon_{1,2}$ associated to $\mathbb{T}_{1,2}$ will give rise to two sets of ward identity for $X(u_{1},v_{1},u_{2},v_{2})$. Ward identities associated to $\mathbb{T}_{1}$ are\footnote{Here SCT refers to special conformal transformation.}
\begin{align}
&\textbf{Translation:} \; \left( \partial_{u_{1}} + \partial_{v_{2}} \right)X(u_{1},v_{1},u_{2},v_{2}) = 0 \\
&\textbf{Scaling:} \; \left( u_{1}\partial_{u_{1}} +h_{1}+ v_{2}\partial_{v_{2}}+ \bar{h}_{2} \right)X(u_{1},v_{1},u_{2},v_{2}) = 0 \\
&\textbf{SCT:} \; \left( u_{1}^{2}\partial_{u_{1}} +2h_{1}u_{1}+ v_{2}^{2}\partial_{v_{2}}+ 2\bar{h}_{2}v_{2} \right)X(u_{1},v_{1},u_{2},v_{2}) = 0 
\end{align}
Similarly for $\mathbb{T}_{2}$ we have the following Ward identities:
\begin{align}
&\textbf{Translation:} \; \left( \partial_{u_{2}} + \partial_{v_{1}} \right)X(u_{1},v_{1},u_{2},v_{2}) = 0 \\
&\textbf{Scaling:} \; \left( u_{2}\partial_{u_{2}} +h_{2}+ v_{1}\partial_{v_{1}}+ \bar{h}_{1} \right)X(u_{1},v_{1},u_{2},v_{2}) = 0 \\
&\textbf{SCT:} \; \left( u_{2}^{2}\partial_{u_{2}} +2h_{2}u_{2}+ v_{1}^{2}\partial_{v_{1}}+ 2\bar{h}_{1}v_{1} \right)X(u_{1},v_{1},u_{2},v_{2}) = 0 
\end{align}
Using both of these Ward identities, one can easily fix $X$ as
\begin{align}
X(u_{1},v_{1},u_{2},v_{2}) = c' \delta_{h_{1},\bar{h}_{2}}\delta_{h_{2},\bar{h}_{1}}(u_{1}-v_{2})^{-(h_{1}+\bar{h}_{2})} (u_{2}-v_{1})^{-(h_{2}+\bar{h}_{1})}
\end{align}
Where $c'$ is an undetermined constant that can be absorbed in the field redefinition. Using those Ward identities, we can also fix the form of a mixed three-point correlator up to an undetermined structure constant. On the other hand, if one considers purely chiral operators, the mixed two-point correlators can be again fixed by those Ward identities:
\begin{align}\label{holo}
&\langle\mathcal{O}_{L}^{(h_{1},0)}(u)\mathcal{\bar{O}}_{R}^{(0,\bar{h}_{2})}(v)\rangle = \frac{c'}{(u-v)^{h_{1}+\bar{h}_{2}}}\delta_{h_{1},\bar{h}_{2}} \\
& \langle\mathcal{O}_{L}^{(h_{1},0)}(u_{1})\mathcal{O}_{R}^{(h_{2},0)}(u_{2})\rangle = 0
\end{align}\\
\textit{\underline{Virasoro algebra}:}
We define $\langle T_{L} \bar{T}_{R} \rangle \propto c_{L\bar{R}}$ and $\langle \bar{T}_{L} T_{R} \rangle \propto c_{\bar{L}R}$. Then from (\ref{spcl bc}), we can see
\begin{align}
c_{L\bar{R}} = c_{L} = \bar{c}_{R}; \; c_{\bar{L}R} = \bar{c}_{L} = c_{R}.
\end{align}
On the other hand from (\ref{holo}), we can see $\langle T_{L} T_{R}\rangle = \langle \bar{T}_{L} \bar{T}_{R} \rangle = 0$. Let us define the modes of $\mathbb{T}_{1,2}$ as
\begin{align}
&\mathbb{L}_{n}^{1} \equiv \frac{1}{2\pi i}\oint du u^{n+1} \mathbb{T}_{1}(u) \\
&\mathbb{L}_{n}^{2} \equiv \frac{1}{2\pi i}\oint du u^{n+1} \mathbb{T}_{2}(u) 
\end{align}
Using (\ref{T1}), we get the commutator 
\begin{align}
&[\mathbb{L}_{n}^{1},\mathbb{L}_{m}^{1}] \nonumber \\
& = \left(\frac{1}{2\pi i}\right)^{2} \left[ \left(\int_{UHP} du_{1} u_{1}^{n+1} T_{L}(u_{1})+\int_{LHP} dv_{1} v_{1}^{n+1} \bar{T}_{R}(v_{1})\right), \left(\int_{UHP} du_{2} u_{2}^{m+1} T_{L}(u_{2})+\int_{LHP} dv_{2} v_{2}^{m+1} \bar{T}_{R}(v_{2})\right)\right]
\end{align}
Here, the contours $UHP$ and $LHP$ refer to the unit semicircles in the upper half plane and lower half-plane, respectively, with a clockwise direction.
\begin{align}
&[\mathbb{L}_{n}^{1},\mathbb{L}_{m}^{1}] \nonumber \\
& = \left(\frac{1}{2\pi i}\right)^{2} \left[ \int_{UHP} du u_{1}^{n+1} T_{L}(u_{1}), \int_{UHP} du_{2} u_{2}^{m+1} T_{L}(u_{2})\right] + \left(\frac{1}{2\pi i}\right)^{2} \left[\int_{LHP} dv_{1} v_{1}^{n+1} \bar{T}_{R}(v_{1}), \int_{LHP} dv_{2} v_{2}^{m+1} \bar{T}_{R}(v_{2})\right]
\end{align}
Since $T_{L}\bar{T}_{R}$ and $\bar{T}_{L}T_{R}$ OPE has non-zero contribution at the real line $u_{i}=v_{i}$, one can show that commutators involving those terms will be zero\footnote{Since the closed contour integration reduces to real line integration, the two integrals commute trivially.}. Using the $TT$ OPE and the following identity
\begin{align}
\left[ \int_{c} du a(u), \int_{c} d\omega b(\omega)\right] = \int_{c}d\omega \oint_{\omega}du a(u) b(\omega)
\end{align}
we get
\begin{align}\label{comm}
& \left(\frac{1}{2\pi i}\right)^{2} \left[ \int_{UHP} du_{1} u_{1}^{n+1} T_{L}(u_{1}), \int_{UHP} du_{2} u_{2}^{m+1} T_{L}(u_{2})\right] \nonumber \\
& =  \left(\frac{1}{2\pi i}\right)^{2} \int_{UHP}du_{2} u_{2}^{m+1} \oint_{u_{2}}du_{1} u_{1}^{n+1} \left(\frac{c_{L}}{2(u_{1}-u_{2})^{4}}+\frac{2T_{L}(u_{2})}{(u_{1}-u_{2})^{2}}+\frac{\partial T_{L}(u_{2})}{u_{1}-u_{2}}\right) \nonumber \\
& = \frac{1}{2\pi i}\frac{c_{L}}{12}n(n^{2}-1)\int_{UHP}du_{2} u_{2}^{m+n+1} + \frac{(n-m)}{2\pi i}\int_{UHP}du_{2} u_{2}^{m+n+1} T_{L}(u_{2})
\end{align} 
Similarly, we will also get
\begin{align}\label{anti comm}
&\left(\frac{1}{2\pi i}\right)^{2} \left[\int_{LHP} dv_{1} v_{1}^{n+1} \bar{T}_{R}(v_{1}), \int_{LHP} dv_{2} v_{2}^{m+1} \bar{T}_{R}(v_{2})\right] \nonumber \\
& = \frac{1}{2\pi i}\frac{\bar{c}_{R}}{12}n(n^{2}-1)\int_{LHP}dv_{2} v_{2}^{m+n+1} + \frac{(n-m)}{2\pi i}\int_{LHP}dv_{2} v_{2}^{m+n+1} \bar{T}_{R}(v_{2})
\end{align}
Thus combining (\ref{comm}) and (\ref{anti comm}) we have,
\begin{align}
&[\mathbb{L}^{1}_{n},\mathbb{L}^{1}_{m}] \nonumber \\
&=\frac{1}{2\pi i}\left(\frac{c_{L}}{12}n(n^{2}-1)\int_{UHP}du_{2} u_{2}^{m+n+1} + \frac{\bar{c}_{R}}{12}n(n^{2}-1)\int_{LHP}dv_{2} v_{2}^{m+n+1} \right)+ \nonumber \\
&+\frac{(n-m)}{2\pi i}\left(\int_{UHP}du_{2} u_{2}^{m+n+1} T_{L}(u_{2})+\int_{LHP}dv_{2} v_{2}^{m+n+1} \bar{T}_{R}(v_{2})\right)
\end{align}
Since $c_{L}=\bar{c}_{R}$ and in the real line $u_{2}=v_{2}$ , $T_{L}(u_{2}) = \bar{T}_{R}(v_{2})$, the $UHP$ and $LHP$ contour merges to give a closed contour of unit circle i.e. $\int_{IHP} +\int_{LHP} \rightarrow \oint_{c}$. Hence, we finally get
\begin{align}
[\mathbb{L}^{1}_{n},\mathbb{L}^{1}_{m}] = \frac{c_{L}}{12} n(n^{2}-1)\delta_{n+m,0} + (n-m)\mathbb{L}^{1}_{n+m}
\end{align}
In a similar fashion, we will also have,
\begin{align}
[\mathbb{L}^{2}_{n},\mathbb{L}^{2}_{m}] = \frac{c_{R}}{12} n(n^{2}-1)\delta_{n+m,0} + (n-m)\mathbb{L}^{2}_{n+m}
\end{align}
Since the OPE of $T_{L}T_{R}, \bar{T}_{L}\bar{T}_{R}, T_{L}\bar{T}_{L}$ and $T_{R}\bar{T}_{R}$ are vanishing, the commutator $[\mathbb{L}^{1}_{n},\mathbb{L}^{2}_{m}] = 0$. Thus, we end up with two copies of Virasoro algebra.\\\\
\textit{\underline{Boundary state}:}
The boundary condition (\ref{spcl bc}) defines a boundary state
\begin{align}
(T_{L}-\bar{T}_{R})|I_{h}\rangle = 0, \; (\bar{T}_{L}-T_{R})|I'_{h}\rangle = 0
\end{align}
If we expand $T_{L} = \sum_{n}u^{-n-2}L_{n}^{1}$ and $T_{R}=\sum_{n}u^{-n-2}L_{n}^{2}$(similarly for anti-holomorphic part), then the boundary states are defined as\footnote{The standard way to construct this is to map UHP to a disk and use mode expansion there.}
\begin{align}
(L_{n}^{1}-\bar{L}_{-n}^{2})|I_{h}\rangle = 0, \; (L_{n}^{2}-\bar{L}_{-n}^{1})|I'_{h}\rangle = 0
\end{align}
If we take copies of the same CFT then we can construct same orthonormal basis of states $|\vec{k},h\rangle$ and $L_{n}^{1}|\vec{k},h\rangle = L_{n}^{2}|\vec{k},h\rangle$. Here $|\vec{k},h\rangle$ is an orthonormal basis of descendants, and $\vec{k}$ refers to infinite dimensional vectors which represent the weight of the descendants at each level\footnote{Here we follow the notation of \cite{Miyaji:2014mca}.}. Once this is true, we know $|I_{h}\rangle$ and $I'_{h}\rangle$ have a similar form of Ishibashi state.
\begin{align}
|I_{h}\rangle = \sum_{\vec{k}}|\vec{k},h\rangle_{L} \otimes |\bar{\vec{k}},h\rangle_{\bar{R}}, \; |I'_{h}\rangle = \sum_{\vec{k}}|\vec{k},h\rangle_{R} \otimes |\bar{\vec{k}},h\rangle_{\bar{L}}
\end{align}

\bibliographystyle{JHEP}
\bibliography{ICFT}

\end{document}